\documentclass[a4paper,preprint,aps,pra]{revtex4}

\usepackage{graphicx}
\usepackage{dcolumn}
\usepackage{bm}
\usepackage{color}

\newcommand{\eqa}{\begin{equation}}
\newcommand{\eqz}{\end{equation}}
\newcommand{\eqma}{\begin{eqnarray}}
\newcommand{\eqmz}{\end{eqnarray}}

\begin{document}
\newcommand{\e}{{\em e}~}
\title{Atomization energies of the carbon clusters C$_n$ (n=2--10) revisited by means of W4 theory as well as density functional, G$n$, and CBS methods\footnote{Henry F. Schaefer III {\em Festschrift}}}
\author{Amir Karton, Alex Tarnopolsky, and Jan M. L. Martin*}
\affiliation{Department of Organic Chemistry, 
Weizmann Institute of Science, 
IL-76100 Re\d{h}ovot, Israel}
\email{gershom@weizmann.ac.il}

\date{{\em Mol. Phys.} TMPH-2008-0396: Received Nov. 24, 2008, accepted Dec. 19, 2008}

\begin{abstract}
The thermochemistry of the carbon clusters C$_n$ (n=2--10) has been revisited by means of W4 theory and W3.2lite theory. Particularly the larger clusters exhibit very pronounced post-CCSD(T) correlation effects. Despite this, our best calculated total atomization energies agree surprisingly well with 1991 estimates obtained from scaled CCD(ST)/6-31G* data. 
Accurately reproducing the small singlet-triplet splitting in C$_2$ requires inclusion of connected quintuple and sextuple excitations.
Post-CCSD(T) correlation effects in C$_4$ stabilize the linear form.
Linear/cyclic equilibria in C$_6$, C$_8$, and C$_{10}$ are not strongly affected by connected quadruples, but they are affected by higher-order triples, which favor polyacetylenic rings but disfavor cumulenic ones. Near the CCSD(T) basis set limit, C$_{10}$ does undergo bond angle alternation in the bottom-of-the-well structure, although it is expected to be absent in the vibrationally averaged structure.
The thermochemistry of these systems, and particularly the longer linear chains, is a particularly difficult test for density functional methods. Particularly for the smaller chains and the rings, double-hybrid functionals clearly outperform convential DFT functionals for these systems. Among compound thermochemistry schemes, G4 clearly outperforms the other members of the G$n$ family.
Our best estimates for total atomization energies at 0 K are: C$_2$($^1\Sigma_g^+$) 144.07, C$_2$($^3\Pi_u$) 142.39, C$_3$($^1\Sigma_g^+$) 315.83, C$_4$($^3\Sigma_g^-$) 429.16, C$_4$($^1A_g$) 430.09, C$_5$($^1\Sigma_g^+$) 596.64, C$_6$($^3\Sigma_g^-$) 717.19, C$_6$($^1A_1^\prime$) 729.68, C$_7$($^1\Sigma_g^+$) 877.45, C$_8$($^3\Sigma_g^-$) 1001.86, C$_8$($^1A_g$) 1014.97, C$_9$($^1\Sigma_g^+$) 1159.21, C$_{10}$($^3\Sigma_g^-$) 1288.22, and C$_{10}$($^1A_1^\prime$) 1355.54 kcal/mol.
\end{abstract}

\maketitle

\section{Introduction}

Carbon clusters, C$_n$, are on the one hand precursors for diamond films generated by CVD (chemical vapor deposition), and on the other hand the smaller congeners of fullerenes and carbon nanotubes.
Their chemistry has been reviewed, among others, by Van Orden and Saykally\cite{Say98}, by Lifshitz\cite{Lif00}, and most recently in the Introduction to Ref.\cite{Ohn08}.

One of the earliest papers to offer a coherent picture of chemical bonding in the C$_n$ clusters for lower $n$ was the seminal study of Raghavachari and Binkley (RB)\cite{Rag87}. While based on what by present-day standards are fairly primitive calculations (HF/6-31G* optimizations and frequency calculations accompanied by CCD(ST)/6-31G* single-point energy evaluations), RB established a number of key features of the chemistry of these systems. This work was expanded upon to some degree by Hutter, L\"uthi, and Diederich (HLD)\cite{HLD}, who carried out DFT optimizations and frequency calculations through C$_{18}$, and by Martin and Taylor (MT1)\cite{Mar96}, who carried out CCSD(T) coupled cluster optimizations and harmonic frequency calculations with fairly modest basis sets, reviewed the experimental information available at this point, and proposed a number of reassignments of infrared spectroscopic features. Guided by predictions in this latter study, cyclic C$_6$\cite{exptC6} and C$_8$\cite{exptC8} were experimentally discovered in graphite vapor trapped in a solid argon matrix. (The first experimental evidence for nonlinear small carbon clusters were probably Coulomb explosion experiments carried out at the Weizmann Institute\cite{Zaj92}.)

Earlier, linear C$_4$ and C$_5$ were correctly assigned\cite{C4C5expt} following computational predictions\cite{C4C5calcs}, and the characterization of linear C$_6$ likewise involved an interplay between theory\cite{C7C9calc} and experiment\cite{C6expt}. Linear C$_7$ and C$_9$ were discovered\cite{C7C9expt} following computational predictions\cite{C7C9calc}.

As discussed by RB and MT1, the odd-numbered clusters C$_{2n+1}$ have linear structures with closed-shell $^1\Sigma^+_g$ ground states, while for the even-numbered clusters, linear structures in $^3\Sigma^-_g$ ground states are energetically close to closed-shell singlet ring structures. Among the latter, the $C_{4n+2}$ rings are cumulenic with $D_{(2n+1)h}$ symmetry and exhibit twin aromatic systems (one conventional, another in-plane): while the ring strain on the in-plane system in C$_6$ is simply too great and leads to both alternating-angle distortion and the ring being less stable than the chain. C$_{10}$ is the first carbon cluster for which the ring structure is more stable than the chain. As for the C$_{4n}$ clusters, while the diamond-shaped C$_4$ ring is nearly isoenergetic with the chain\cite{Mar96c4,Schwarz2000}, the C$_8$ ring is polyacetylenic and exhibits both bond length and bond angle alternation ($C_{4h}$ symmetry), and larger C$_{4n}$ rings were predicted\cite{Mar95c18} to generally have polyacetylenic $C_{(2n)h}$ structures.
At some point for high $4n+2$, Peierls distortion will set in: this issue has been discussed and reviewed at length in Ref.\cite{Ohn08}.

Thermochemistry was addressed in a number of theoretical and experimental studies. RB applied an empirical scaling factor of 1.1 to their CCD(ST)/6-31G* atomization energies, which they derived from the ratio between their calculated values and Knudsen effusion measurements by Drowart et al.(DBDI)\cite{Dro59} for C$_2$ through C$_5$. Since the DBDI values for C$_4$ and C$_5$ were third-law extrapolations based on very crude estimates of the molecular constants, the scaled estimates of RB were biased upwards, and the scaling factor required downward revision to 1.082\cite{Mar91}. 
Later, new measurements by Gingerich and coworkers\cite{Ging94} became available, which also covered C$_6$ and C$_7$. Subsequently, Martin and Taylor (MT2)\cite{Mart95} proposed revised estimates based on CCSD(T)/cc-pVTZ and CCSD(T)/cc-pVQZ calculations with empirical basis set incompleteness corrections. 

Both computational chemistry in general, and computational thermochemistry in particular, have come a long way since this time. In particular, our group recently developed a post-CCSD(T) computational thermochemistry protocol known as W4 theory\cite{W4}, which exhibits an RMSD deviation from experiment (Active Thermochemical Tables, ATcT\cite{ATcT}, values) of just 0.08 kcal/mol, implying a 95\% confidence interval of just 0.16 kcal/mol. The purpose of the present paper is to furnish the best thermochemistry for the lower C$_n$ clusters feasible with current technology.

Carbon clusters are among the manifold research interests of Prof. Henry F. Schaefer III\cite{Liang90,SchaeferC10,Grev90,SchaeferC2,Schaefer2007}, who is being honored by the present issue. 

\section{Computational methods}
The self consistent field (SCF), ROCCSD and ROCCSD(T) calculations\cite{Rag89Wat93} were carried out using version 2006.1 of the Molpro program system\cite{molpro}. All single-point post-CCSD(T) calculations were carried out using an OpenMP-parallel version of Mih\'aly K\'allay's general coupled cluster code MRCC\cite{mrcc} interfaced to the Austin-Mainz-Budapest version of the ACES II program system\cite{aces2de}. The diagonal Born-Oppenheimer correction (DBOC) calculations were carried out using the Austin-Mainz-Budapest version of the ACES II program system as well as using PSI3\cite{psi3}.

For the lower-cost methods W2.2 and W3.2lite we used reference geometries optimized at the B3LYP/cc-pVTZ level of theory, while for the more rigorous methods W3.2, W4lite, and W4 we used CCSD(T)/cc-pVQZ reference geometries. The optimized geometries are summarized in Table \ref{tab:geom}.
 
For the large scale SCF, CCSD, CCSD(T), and post-CCSD(T) single point calculations, we employed the cc-pV{\em n}Z basis set\cite{Dun89}. In core-valence correlation calculations, the core-valence weighted correlation consistent basis sets of Peterson and Dunning were employed\cite{pwCVnZ}. Scalar relativistic calculations were carried out using the Pacific Northwest National Laboratory (PNNL) Douglas-Kroll-Hess relativistically contracted correlation basis sets\cite{Dyall}. 

The ROHF-SCF contribution is extrapolated using the Karton-Martin\cite{Kar-Mar} modification of Jensen's extrapolation formula.\cite{Jen05} All other extrapolations are carried out using the $A+B/L^\alpha$ two-point extrapolation formula (where $L$ is the highest angular momentum present in the basis set). The ROCCSD valence contribution is partitioned into singlet-pair energies, triplet-pair energies and $\hat{T}_1$ terms.\cite{Klop01} The singlet- and triplet-pair energies are extrapolated with $\alpha_S=$3 and $\alpha_T=$5, respectively, while the $\hat{T}_1$ term (which exhibits very weak basis set dependence) is simply set equal to that in the largest basis set. All other extrapolations are carried out with $\alpha=$3.\cite{W4,W4.4}

The W{\em n} family of methods W1, W2.2, W3.2lite, W3.2, W4lite, and W4 used in the present study provide a sequence of converging computational thermochemistry protocols. A detailed description and rationalization of the W{\em n} protocols is given elsewhere.\cite{W2,W1h,W3,W4,W3.2lite} 
For the purpose of the present paper we use the W{\em n}h variants of the W{\em n} methods, in which the diffuse functions are omitted from carbon and less electronegative elements\cite{W1h}. (For the present all-carbon systems, this approximation is of no thermochemical consequence,
but computer resource requirements are substantially reduced.) 
In W4h theory the SCF and valence CCSD contributions to the TAE are extrapolated from cc-pV5Z and cc-pV6Z basis sets, and the valence parenthetical triples (T) contribution from cc-pVQZ and cc-pV5Z basis sets. The higher-order connected triples, $\hat{T}_3-$(T), valence correlation contribution is extrapolated from the cc-pVDZ and cc-pVTZ basis sets. As for the connected quadruple, $\hat{T}_4$, term---the (Q) and $T_4-$(Q) corrections are calculated with the cc-pVTZ and cc-pVDZ basis sets, respectively, both scaled by 1.1. This formula offers a very reliable as well as fairly cost-effective estimate of the basis set limit $\hat{T}_4$ contribution.\cite{W4,W4.4} The $\hat{T}_5$ contribution is calculated using the $sp$ part of the cc-pVDZ basis set, denoted cc-pVDZ(no d). The CCSD(T) inner-shell contributions are extrapolated from cc-pwCV{\em n}Z basis sets ($n=$T, Q). 
Scalar relativistic contributions (second-order Douglas-Kroll-Hess approximation\cite{DKH1DKH2}) are obtained from the difference between nonrelativistic CCSD(T)/cc-pVQZ and CCSD(T)/cc-pVQZ-DK calculations. Atomic and molecular first-order spin-orbit coupling terms are taken from the experimental fine structure. Finally, the DBOC is calculated at the ROHF/cc-pVTZ level of theory. 

When considering the W$n$ sequence, it may be helpful to keep the following points in mind: (a) W1 and W2.2 completely neglect post-CCSD(T) correlation effects; (b) the difference between W1 and W2.2 is (in the context of this paper) principally a matter of more reliable CCSD(T) basis set limits and the DBOC; (c) the only differences between W2.2 on the one hand and W3.2lite and W3.2 on the other hand are post-CCSD(T) corrections, the former with a cost-saving empirical approximation;
(d) the only difference between W3.2 and W4lite is a more reliable basis set limit for the CCSD(T) part; (e) the only improvement in W4 relative to W4lite is a more rigorous account for connected quadruple and higher excitations.

All density functional results were obtained by means of a locally modified version of Gaussian 03\cite{g03}.

\section{Results and Discussion}

Diagnostics for the importance of nondynamical correlation  can be found in Table \ref{tab:diagnostics}. These include the ${\cal T}_1$ and ${\cal D}_1$ diagnostics\cite{Lee89,Lei00}, 
the largest $T_2$ amplitudes at the CCSD/cc-pVTZ level, the HOMO and LUMO natural orbital occupations at the same level, and percentages of the total atomization energy accounted for at the SCF level, by parenthetical triples (shown elsewhere\cite{W4} to be a more reliable predictor of the importance of post-CCSD(T) correlation effects than any other diagnostic), and by post-CCSD(T) correlation effects.

Table \ref{tab:W3.2L} presents component breakdowns for the W3.2lite results, Table \ref{tab:W4} for the W4 results (C$_2$ through C$_5$ only), and Table \ref{tab:TAE0} a comparison between theoretical and experimental estimates for C$_2$ through C$_{10}$.

For C$_2$, a well-known `problem molecule',\cite{SchaeferC2,C2,C2Bartlett} 
an ATcT value is available:\cite{Feller} our W4 calculation agrees with it to within overlapping uncertainties. Still higher-level W4.3 and W4.4 calculated values are in basically perfect agreement with the ATcT value.\cite{W4,W4.4} 
At the W4 level, the calculated singlet-triplet splitting is found to be 609 cm$^{-1}$, somewhat lower than the experimental value\cite{Hub79} of 716 cm$^{-1}$ (2.05 kcal/mol). Increasing the level of theory to W4.2 has a negligible effect on the triplet state (which has much 
weaker multireference character) but stabilizes the singlet state by a small but significant amount. As a result, at the W4.2 level
the splitting goes up to 662 cm$^{-1}$  --- the difference with W4 is
entirely due to higher-order $T_3$ effects in the core-valence term. Further `ramping up' to the W4.3 level sees a slight destabilization of the triplet state (mostly due to the $T_3-(T)$ and $T_4$ contributions being calculated with larger basis sets), but a somewhat more pronounced stabilization of the singlet state due to quintuple and sextuple excitations. Consequently, perfect agreement with experiment is
reached, with an energy difference of 2.05 kcal/mol, or 2.00 kcal/mol if the experimental spin-orbit coupling constant\cite{Hub79}, $A$=-15.25 cm$^{-1}$,
is taken into account. (It is not entirely clear whether the experimental $T_e$=716 cm$^{-1}$ refers to the lowest spin-orbit component of the triplet state or to the average of the three spin-orbit components. Incidentally, our calculated $A=-14.39$ cm$^{-1}$ using the same procedure as in Ref.\cite{BN}.) 
 For the isoelectronic BN diatomic --- which exhibits even more pathological nondynamical correlation effects --- we likewise found\cite{BN} that reproduction of the very small singlet-triplet splitting required accounting for connected sextuple and especially quintuple excitations, contributions of which have similar magnitudes as in C$_2$. For the sake of completeness, our computed $T_e$ for the $b~^3\Sigma^-_g$ state is 18.16 kcal/mol at the W4 level and 18.44 kcal/mol at the W4.3 level, compared to an experimental value\cite{Hub79} of 18.40 kcal/mol. Once again, essentially all post-W4 change is due to the pathological singlet state.

\begin{figure}
\includegraphics[width=12cm,angle=0]{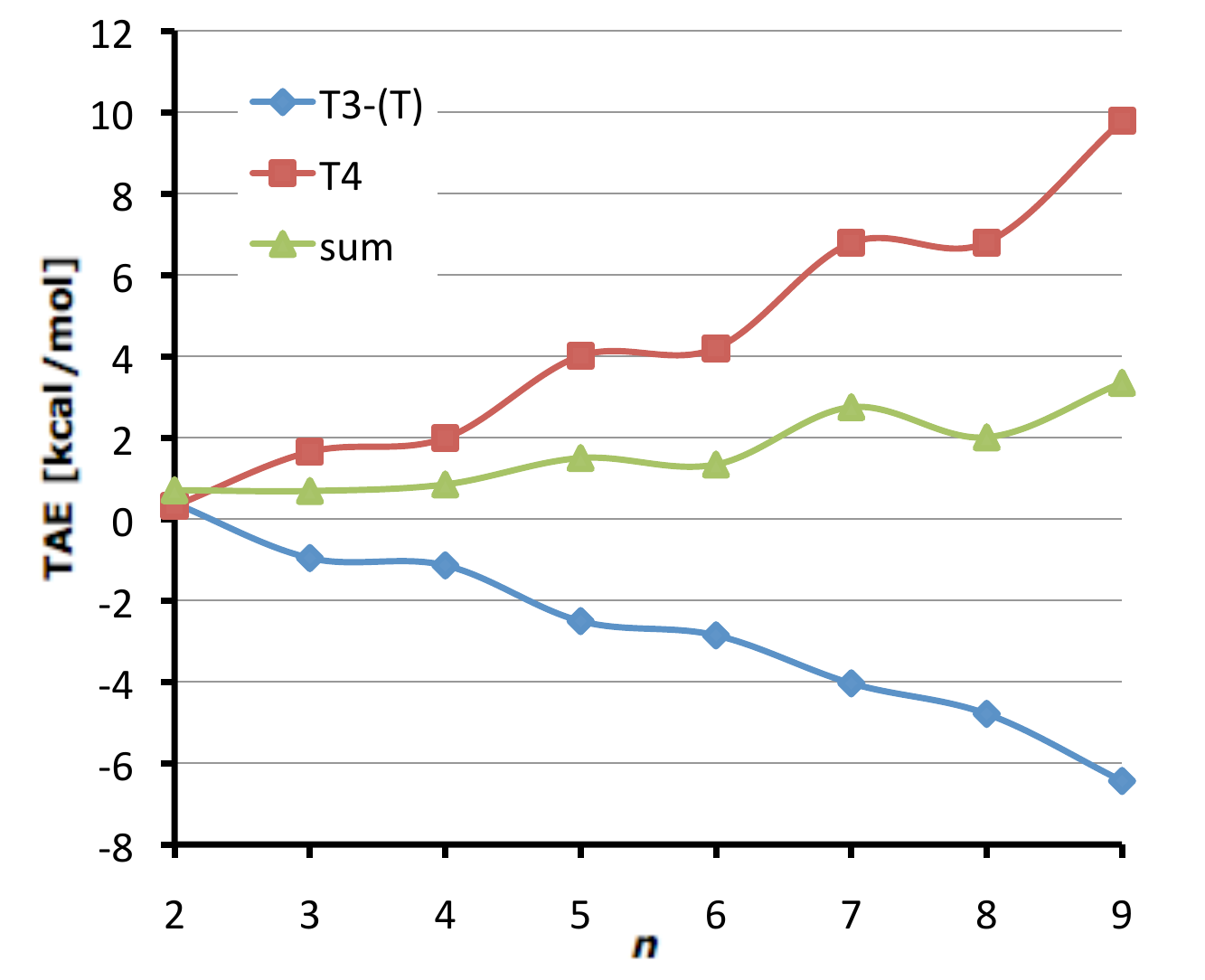} \\
\caption{$\hat{T}_3-$(T) and $\hat{T}_4$ contributions to the total atomization energies for the linear C$_n$ clusters (from W3.2lite theory, in kcal/mol)\label{fig:figure1}}
\end{figure}

Perusing both Tables \ref{tab:W3.2L} and \ref{tab:W4}, it becomes evident that higher-order correlation effects wax increasingly more prominent as $n$ goes up. 
This phenomenon is illustrated graphically in Figure~\ref{fig:figure1}.
For example, higher-order triples for C$_9$ reduce TAE by 6.4 kcal/mol, while connected quadruples will increase it by approximately 9.8 kcal/mol. While these two effects partially compensate each other, together this still amounts to an n-particle truncation error at the CCSD(T) level of about 3.4 kcal/mol. For comparison, the same effect is just 0.7 kcal/mol (W3.2lite) or 1.0 kcal/mol (W4) for C$_3$, and 0.8 kcal/mol (W3.2lite) for C$_6$. 

Connected quintuples likewise increase in importance as $n$ grows, reaching nearly 1 kJ/mol for C$_5$ (the largest system for which we were able to calculate their contribution explicitly).

Interestingly enough, with the anomalous exception of C$_2$($^1\Sigma_g^+$), all molecules considered here have \%TAE[(T)] diagnostics hovering in the 5\% range, which indicates moderate but not severe nondynamical correlation\cite{W4}.

We also note that the very small linear-cyclic energy difference in C$_4$ has an appreciable post-CCSD(T) contribution: at the W4 level, higher-order connected triples and connected quadruples favor the linear structure by 0.5 and 0.4 kcal/mol, respectively. Likewise, in C$_6$, higher-order triples favor the linear structure by 0.5 kcal/mol (W3.2) while connected quadruples only affect the equilibrium by 0.08 kcal/mol (favoring the ring). In polyacetylenic C$_8$, however, higher-order triples favor the ring by 2 kcal/mol, while the effect of connected quadruples is nearly an order of magnitude smaller. This raises the question whether a general trend exists: higher-order triples favoring polyacetylenic rings over linear structures while favoring linear over cumulenic ring structures. Unfortunately, CCSDT calculations beyond C$_8$ are beyond our computational resources.

The equilibrium structure of the C$_{10}$ ring displays clear angle alternation. It was previously noted\cite{Mar96} that the barrier towards the $D_{10h}$ saddle point is very small, and there has been some speculation (e.g., Ref.\cite{SchaeferC10,Watts92,Parasuk92,QMC})
that it might actually disappear at the basis set limit. Our calculated CCSD(T)/cc-pVTZ, CCSD(T)/cc-pVQZ(no g), and CCSD(T)/cc-pVQZ barriers are 1.11, 0.48, and 0.33 kcal/mol, respectively.
At the \{Q,5\} basis set limit at  CCSD(T)/cc-pVQZ optimum geometries, the $D_{5h}$ alternating-angles ring is 23.00 kcal/mol more stable at the SCF level than the $D_{10h}$ saddle point. CCSD valence correlation lowers this by 18.37 kcal/mol to 4.63 kcal/mol, 
while parenthetical triples shave off yet another 4.32 kcal/mol. These components add up to a basis set limit CCSD(T) deformation energy of 0.30 kcal/mol. MP2, as expected\cite{Mar95c18,Mar96}, grossly favors the
more symmetric structure,
and SCS-MP2/cc-pVQZ\cite{SCS-MP2} is unable to overcome this bias, reaching
-26.75 kcal/mol for the isomerization energy. At the SCS-CCSD/cc-pVQZ\cite{SCS-CCSD} level, however, +0.15 kcal/mol is obtained,
slightly increasing to 0.18 kcal/mol at the SCS-CCSD/cc-pV5Z//CCSD(T)/cc-pVQZ level. These numbers are as close to the CCSD(T)
answer as one can reasonably hope. 

Notwithstanding the above, the calculated B3LYP/cc-pVTZ zero-point vibrational energy of the saddle point is about 1 kcal/mol smaller than for the D$_{5h}$ minimum. This implies that the vibrationally averaged structure will have $D_{10h}$ symmetry even at 0 K.

Inner-shell correlation contributions become fairly hefty for the larger clusters, reaching 12.6 kcal/mol for C$_{10}$. For the linear C$_n$ clusters, these contributions scale almost perfectly linearly with $n$. For C$_6$, C$_8$, and C$_{10}$, inner-shell correlation systematically favors the cyclic structures: the somewhat anomalous bicyclic C$_4$ structure does not follow this trend.

It was earlier noted by RB, for the linear clusters, that the reactions 2C$_n$$\rightarrow$C$_{n+2}$+C$_{n-2}$ are nearly thermoneutral at their level of theory. We do find the same to be the case at the W3.2lite level: however, relative to the (quite small) overall reaction energies, the post-CCSD(T) contributions are still non-negligible. 
If we assume that those reaction energies are converged at the W3.2lite level (which is probably justified to 0.5 kcal/mol, and certainly to 1 kcal/mol), we can obtain extrapolated W4 atomization energies for species heavier than C$_5$ (the practical limit for full W4 calculations with the presently available hardware). Our best estimates are presented in Table \ref{tab:TAE0}.

Perhaps the most striking feature of these numbers is the outstanding agreement between our best calculated TAE$_0$ values for the linear clusters and the earlier, empirically scaled, data of MFG. Indeed, it is hard to believe that these latter rudimentary estimates -- blissfully ignorant of higher-order correlation effects, inner-shell correlation, etc. --- capture so much of the thermochemistry in these systems. We note that this scaling is not transferable to the rings, for which the MFG estimates are clearly too low.

Our best values agree reasonably well with the available measured data of Gingerich and coworkers\cite{Ging94}, considering the size of the uncertainties on the latter --- although generally the calculated values are near the upper edges of the experimental error bars.

Inspection of Table III also suggests (not surprisingly) that as $n$ increases, the C$_n$ atomization energies at the W3.2lite level are progressively greater overestimates compared to the estimated W4 numbers. 

In conjunction with the ATcT revised heat of formation of carbon atom\cite{ATcT}$^{(c)}$, 
$\Delta H^\circ_{f,0}$[C($g$)]=170.06$\pm$0.026 kcal/mol,
we can offer the following revised heats of formation at 0 K: C$_2$($^1\Sigma_g^+$) 196.04, C$_2$($^3\Pi_u$) 197.72, C$_3$($^1\Sigma_g^+$) 194.33, C$_4$($^3\Sigma_g^-$) 251.06, C$_4$($^1A_g$) 250.13, C$_5$($^1\Sigma_g^+$) 253.63, C$_6$($^3\Sigma_g^-$) 304.14, C$_6$($^1A_1^\prime$) 290.65, C$_7$($^1\Sigma_g^+$) 312.94, C$_8$($^3\Sigma_g^-$) 358.58, C$_8$($^1A_g$) 345.47, C$_9$($^1\Sigma_g^+$) 371.29, C$_{10}$($^3\Sigma_g^-$) 412.33, and C$_{10}$($^1A_1^\prime$) 345.01 kcal/mol.

Table \ref{tab:Gn} presents computed atomization energies, and error statistics relative to our best values, for more approximate compound thermochemistry methods such as G2MP2\cite{g2mp2}, G2\cite{g2}, G3\cite{g3}, G3B3\cite{g3b3}, G4\cite{g4}, G4(MP2)\cite{G4MP2}, CBS-QB3\cite{cbs-qb3}, CBS-APNO\cite{cbs-apno}, W1h\cite{W2}, and W2.2 for the thermochemistry of these systems.

Excluding C$_8$ and larger clusters (for which we deem our best calculated values to be less reliable), one sees similar error statistics for W1h and for W2.2, suggesting that an RMSD of about 1 kcal/mol is the best one can hope for without including post-CCSD(T) correlation effects. CBS-APNO and W1h have comparable RMSD errors: however, while the W1h values tend to be underestimates (consistent with the positive post-CCSD(T) correlation effects), the CBS-APNO values tend to be overestimates. The next best performer, G4, errs on both sides.
Somewhat counterintuitively, G2MP2 displays marginally better error statistics than G2 --- despite MP2 being clearly inappropriate for these systems --- and G3 noticeably better ones than G3B3 --- despite several qualitatively wrong reference geometries and spurious imaginary frequencies at the levels of theory used in G3 for reference geometries (MP2/6-31G*) and ZPVEs (HF/6-31G*), respectively. 
We believe the apparently better performance of G3 to be due to error compensation. CBS-QB3 clearly outperforms G3B3: in addition, it errs on the `right' side --- underestimating the best values, consistent with CCSD(T) limits. The unrealistically small linear-cyclic difference calculated for C$_8$ is something of a `clinker' on the part of CBS-QB3, however. One statement we can confidently make is that, for these systems too, G4 clearly outperforms the other members  of the G$n$ series. Performance of G4(MP2), which actually does include a CCSD(T)/6-31G(d) step but considers basis set extension effects at the MP2 level only, is considerably degraded relative to G4, rather more so than generally seen\cite{G4MP2}. This is not surprising in light of the poor performance of MP2 for the present systems.

Finally, we compare the relative performance of different DFT exchange-correlation functionals in predicting the binding energies of the carbon clusters considered in the present work. As reference data we use our best available\cite{DFT_ref_val} nonrelativistic, clamped-nuclei, zero-point exclusive TAEs. The exchange-correlation functionals employed include the following classes:
(a) generalized gradient approximation (GGA): HCTH407,\cite{HCTH407} BLYP,\cite{B88ex,LYPc} BPW91,\cite{B88ex,PW91c} BP86\cite{B88ex,P86c} and PBE\cite{PBE}; (b) meta-GGA: M06-L,\cite{M06-L} VSXC,\cite{VSXC} TPSS\cite{TPSS} and $\tau$-HCTH;\cite{t-HCTH} (c) hybrid GGA: PBE0,\cite{PBE0} B97-2,\cite{B97-2} B3PW91,\cite{B3,PW91c} B97-1,\cite{B97-1} B98,\cite{B98} TPSSh,\cite{TPSSh} B3LYP,\cite{LYPc,B3,B3LYP} mPW1K\cite{mPWex,mPW1K} and BHLYP;\cite{BHLYP} (d) hybrid meta-GGA: mPW1B95,\cite{mPWex,mPW1B95} B1B95,\cite{B88ex,B95c} M06,\cite{M06} PW6B95,\cite{PW6B95} TPSS1KCIS,\cite{TPSS1KCIS} M05,\cite{M05} $\tau$-HCTHh,\cite{t-HCTH} BMK,\cite{BMK} M06-2X,\cite{M06}  BB1K\cite{BB1K} and PWB6K;\cite{PW6B95} and (e) double hybrid functionals: B2-PLYP,\cite{B2-PLYP} mPW2-PLYP,\cite{mPW2-PLYP} B2T-PLYP,\cite{B2K-PLYP} B2K-PLYP\cite{B2K-PLYP} and B2GP-PLYP.\cite{B2GP-PLYP} Unless otherwise indicated, the pc-2 basis set of Jensen\cite{pc-2}, which is of $[4s3p2d1f]$ quality but optimized for Hartree-Fock and DFT calculations, was employed throughout. 
The root mean square deviations (RMSD), mean signed deviations (MSD), and mean average deviations (MAD) are gathered in Table \ref{tab:DFT}. 

In general, the GGA, meta-GGA, hybrid-GGA, and hybrid meta GGA functionals systematically overestimate the binding energies and lead to unacceptable RMSDs of 11--57 kcal/mol. While the so called `kinetics' functionals (HF exchange $>$ 40\%) BHLYP, mPW1K, PWB6K, BB1K, and to a lesser extent BMK tend to systematically underestimate the binding energies. 

Not quite surprisingly, all conventional DFT functionals fare rather poorly, with RMSDs in the double digits overall. Simply eliminating the pathological C$_2$ singlet case brings RMSD below 10 kcal/mol for four functionals: M05, BB1K, M06-2X, and BMK (9.4, 8.4, 7.5, and 6.2 kcal/mol, respectively). 
Interestingly, once the linear systems larger than C$_4$ (and the C$_{10}$ ring) are also eliminated, RMSDs drop sharply almost across the board. Long cumulenic chains are expected to be a `torture test' for any DFT method.

This becomes especially clear for the double hybrids. Near the basis set limit, RMSD for the reduced set of systems actually reaches the 2 kcal/mol range that can generally be expected for functionals like B2GP-PLYP\cite{B2GP-PLYP}. For the long chains, however, severe overestimates set in. We note that these systems have increasingly narrower HOMO-LUMO gaps, and that\cite{Francois} they have exceptionally large electron affinities for large $n$.
It is perhaps not overly surprising that systems that put even post-CCSD(T) methods severely to the test (and, for instance, have connected quadruples contributions that climb to 10 kcal/mol for C$_9$) would be beyond the reach of double hybrids, let alone conventional DFT functionals.

In terms of singlet-triplet equilibria, as noted earlier,\cite{Mar96c11} shortcomings of conventional DFT functionals (and particularly `kinetics' functionals) are exacerbated by a natural bias of hybrid functionals towards high-spin states. This issue is mitigated by the MP2-like correlation in double hybrids.

\section{Conclusions}
The thermochemistry of the carbon clusters C$_n$ (n=2--10) has been revisited by means of W4 theory and W3.2lite theory. Particularly the larger clusters exhibit very pronounced post-CCSD(T) correlation effects. Despite this, our best calculated total atomization energies agree surprisingly well with 1991 estimates obtained from scaled CCD(ST)/6-31G* data. 
Accurately reproducing the small singlet-triplet splitting in C$_2$ requires inclusion of connected quintuple and sextuple excitations.
Post-CCSD(T) correlation effects in C$_4$ stabilize the linear form.
Linear/cyclic equilibria in C$_6$, C$_8$, and C$_{10}$ are not strongly affected by connected quadruples, but they are affected by higher-order triples, which favor polyacetylenic rings but disfavor cumulenic ones. Near the basis set limit, C$_{10}$ does undergo bond angle alternation in the bottom-of-the-well structure, although it is expected to be absent in the vibrationally averaged structure.
The thermochemistry of these systems, and particularly the longer linear chains, is a particularly difficult test for density functional methods. Particularly for the smaller chains and the rings, double-hybrid functionals clearly outperform convential DFT functionals for these systems. Among compound thermochemistry schemes, G4 clearly outperforms the other members of the G$n$ family.
Our best estimates for total atomization energies at 0 K are: C$_2$($^1\Sigma_g^+$) 144.07, C$_2$($^3\Pi_u$) 142.39, C$_3$($^1\Sigma_g^+$) 315.83, C$_4$($^3\Sigma_g^-$) 429.16, C$_4$($^1A_g$) 430.09, C$_5$($^1\Sigma_g^+$) 596.64, C$_6$($^3\Sigma_g^-$) 717.19, C$_6$($^1A_1^\prime$) 729.68, C$_7$($^1\Sigma_g^+$) 877.45, C$_8$($^3\Sigma_g^-$) 1001.86, C$_8$($^1A_g$) 1014.97, C$_9$($^1\Sigma_g^+$) 1159.21, C$_{10}$($^3\Sigma_g^-$) 1288.22, and C$_{10}$($^1A_1^\prime$) 1355.54 kcal/mol.
\section*{Acknowledgments}
Research at Weizmann was funded by the Israel Science Foundation (grant 709/05), the Minerva Foundation (Munich, Germany), and the Helen and Martin Kimmel Center for Molecular Design. JMLM is the incumbent of the Baroness Thatcher Professorial Chair of Chemistry and a member {\em ad personam} of the Lise Meitner-Minerva Center for Computational Quantum Chemistry.

\clearpage

\squeezetable
\begin{table}
\caption{Theoretical geometries (in {\AA}ngstrom and degree). In the linear systems, bond lengths are numbered from the outside in.\label{tab:geom}}
\begin{tabular}{l|ll|cccccccc}
\hline\hline
& & & $r_1$ & $r_2$ & $r_3$ & $r_4$ & $r_5$ & $\alpha$ \\
\hline
             &C$_2$($^1\Sigma_g^+$)   & D$_{\infty h}$& 1.247 &  &  &  &  &  \\
             &C$_2$($^3\Pi_u$)        & D$_{\infty h}$& 1.302 &  &  &  &  &  \\ 
             &C$_2$($^3\Sigma^-_g$)        & D$_{\infty h}$& 1.367 &  &  &  &  &  \\ 
             &C$_3$($^1\Sigma_g^+$)   & D$_{\infty h}$& 1.288 &  &  &  &  &  \\
             &C$_4$($^3\Sigma_g^-$)   & D$_{\infty h}$& 1.306 & 1.287 &  &  &  &  \\ 
             &C$_5$($^1\Sigma_g^+$)   & D$_{\infty h}$& 1.282 & 1.279 &  &  &  &  \\ 
             &C$_6$($^3\Sigma_g^-$)   & D$_{\infty h}$& 1.296 & 1.284 & 1.271 &  &  &  \\ 
B3LYP/pc-2   &C$_7$($^1\Sigma_g^+$)   & D$_{\infty h}$& 1.282 & 1.284 & 1.269 &  &  &  \\ 
             &C$_8$($^3\Sigma_g^-$)   & D$_{\infty h}$& 1.291 & 1.286 & 1.270 & 1.277 &  &  \\ 
             &C$_9$($^1\Sigma_g^+$)   & D$_{\infty h}$& 1.281 & 1.286 & 1.267 & 1.273 &  &  \\ 
             &C$_{10}$($^3\Sigma_g^-$)& D$_{\infty h}$& 1.288 & 1.287 & 1.269 & 1.277 & 1.270 &  \\
             &C$_4$($^1A_g$)          & D$_{2 h}$     & 1.442 &  &  &  &  & 117.6\\ 
             &C$_6$($^1A_1^\prime$)   & C$_{3 h}$     & 1.319 &  &  &  &  & 147.9 \\ 
             &C$_8$($^1A_g$)          & D$_{4 h}$     & 1.252 & 1.380 &  &  &  & 162.1 \\ 
             &C$_{10}$($^1A_1^\prime$)& D$_{5 h}$     & 1.288 &  &  &  &  & 161.8 \\ 
\hline
             &C$_2$($^1\Sigma_g^+$)   & D$_{\infty h}$& 1.2400$^a$ &  &  &  &  &  \\
             &C$_2$($^3\Pi_u$)        & D$_{\infty h}$& 1.3153 &  &  &  &  &  \\ 
             &C$_2$($^3\Sigma^-_g$)        & D$_{\infty h}$& 1.3728 &  &  &  &  &  \\ 
             &C$_3$($^1\Sigma_g^+$)   & D$_{\infty h}$& 1.2981 &  &  &  &  &  \\
             &C$_4$($^3\Sigma_g^-$)   & D$_{\infty h}$& 1.3140 & 1.2936 &  &  &  &  \\ 
             &C$_5$($^1\Sigma_g^+$)   & D$_{\infty h}$& 1.2936 & 1.2857 &  &  &  &  \\ 
             &C$_6$($^3\Sigma_g^-$)   & D$_{\infty h}$& 1.3052 & 1.2905 & 1.2780 &  &  &  \\ 
CCSD(T)/cc-pVQZ&C$_7$($^1\Sigma_g^+$)   & D$_{\infty h}$& 1.2933 & 1.2901 & 1.2759 &  &  &  \\ 
             &C$_8$($^3\Sigma_g^-$)   & D$_{\infty h}$& 1.3010 & 1.2923 & 1.2767 & 1.2832 &  &  \\ 
             &C$_9$($^1\Sigma_g^+$)   & D$_{\infty h}$& 1.2930 & 1.2928 & 1.2745 & 1.2799 &  &  \\  
             &C$_4$($^1A_g$)          & D$_{2 h}$     & 1.4492 &  &  &  &  & 117.09\\ 
             &C$_6$($^1A_1^\prime$)   & D$_{3 h}$     & 1.3282 &  &  &  &  & 148.79 \\ 
             &C$_8$($^1A_g$)          & C$_{4 h}$     & 1.2592 & 1.3926 &  &  &  & 162.74 \\ 
             &C$_{10}$($^1A_1^\prime$)& D$_{5 h}$     & 1.2940 &  &  &  &  &  158.25\\
             &C$_{10}$($^1A_1^\prime$)& D$_{10 h}$    & 1.2914 &  &  &  &  &  [144.00]\\
\hline\hline
\end{tabular}

(a) Fixed reference geometry used for consistency with earlier post-W4 work\cite{W4.4}. Actual CCSD(T)/cc-pVQZ bond distance is 1.2458 \AA.
W4 TAE$_e$ at that geometry is 143.87 kcal/mol, just 0.01 kcal/mol higher than at 1.24 \AA.
\end{table}

\clearpage
\squeezetable
\begin{table}
\caption{Diagnostics for importance of nondynamical correlation\label{tab:diagnostics}}
\begin{tabular}{l|cccc|ccc|cc}
\hline\hline
 & \%TAE$_e$ & \%TAE$_e$ & \%TAE$_e$ & \%TAE$_e$ & ${\cal T}_1$ & $D_1$ & Largest T$_2$ & \multicolumn{2}{c}{NO occupations}\\
 & [SCF]$^a$ & [(T)]$^a$ & [post-CCSD(T)]$^a$ & [$T_4+T_5$]$^a$ & \multicolumn{2}{c}{diagnostic} & amplitudes & HDOMO$^b$ & LUMO\\
 & & & & & \multicolumn{5}{c}{--- CCSD/cc-pVTZ ---} \\
\hline
C$_2$($^1\Sigma_g^+$)              & 12.61 & 13.29 & 0.28 & 1.79 & 0.038 & 0.086 & 0.29 & 1.629 & 0.362\\
C$_2$($^3\Pi_u$)                   & 50.56 & 6.51 & 0.85 & 1.02  & 0.020 & 0.039 & 0.12 & 1.934 & 0.084\\
C$_2$($^3\Sigma_g^-$)              & 67.70 & 3.40 & 0.53 & 0.34  & 0.011 & 0.021 & 0.08 ($\times$2)     & 1.933 & 0.044\\
C$_3$($^1\Sigma_g^+$)              & 64.88 & 5.52 & 0.30 & 0.64  & 0.023 & 0.052 & 0.10 ($\times$2)&1.913 ($\times$2) & 0.074 ($\times$2)\\
C$_4$($^3\Sigma_g^-$)              & 66.09 & 4.61 & 0.30 & 0.62  & 0.018 & 0.035 & 0.07 ($\times$2)& 1.915 ($\times$2) & 0.068 ($\times$2)\\
C$_4$($^1A_g$)                     & 64.44 & 4.70 & 0.10 & 0.54  & 0.015 & 0.034 & 0.09 & 1.908 & 0.074\\
C$_5$($^1\Sigma_g^+$)              & 66.64 & 5.37 & 0.14 & 0.60  & 0.024 & 0.060 & 0.09 ($\times$2)& 1.903 ($\times$2) & 0.084 ($\times$2)\\
C$_6$($^3\Sigma_g^-$)              & 66.73 & 4.79 & 0.10 & 0.54  & 0.023 & 0.050 & 0.08 & 1.912 ($\times$2) & 0.074 ($\times$2)\\
C$_6$($^1A_1^\prime$)              & 64.19 & 5.00 & 0.03 & 0.52  & 0.035 & 0.116 & 0.07 ($\times$2)&1.915 ($\times$2) & 0.098\\
C$_7$($^1\Sigma_g^+$)              & 67.18 & 5.30 & 0.19 & 0.71  & 0.025 & 0.066 & 0.09 ($\times$2)&1.897 ($\times$2) & 0.090 ($\times$2)\\
C$_8$($^3\Sigma_g^-$)              & 67.22 & 4.80 & 0.20 & 0.67  & 0.025 & 0.061 & 0.08 & 1.904 ($\times$2) & 0.0866 ($\times$2)\\
C$_8$($^1A_g$)                     & 66.15 & 4.46 & 0.43 & 0.69  & 0.039 & 0.130 & 0.06 ($\times$2)& 1.889$^d$ &  0.100$^d$\\
C$_9$($^1\Sigma_g^+$)              & 67.59 & 5.18 & 0.29 & 0.83  & 0.025 & 0.071 & 0.08 ($\times$2)& 1.895 ($\times$2) & 0.096 ($\times$2)\\
C$_{10}$($^3\Sigma_g^-$)           & 67.58 & 4.87 & N/A & N/A    & 0.026 & 0.071 & 0.05 ($\times$2) & 1.901 ($\times$2) & 0.907 ($\times$2)\\
C$_{10}$($^1A_1^\prime$)D$_{5 h}$  & 67.95 & 4.68 & N/A & N/A    & 0.037 & 0.112 & 0.07 ($\times$2)& 1.900 ($\times$2)$^{c,d}$ & 0.089 ($\times$2)$^{c,d}$\\
\hline\hline
\end{tabular}
\begin{flushleft}
$^a$Percentages of the total atomization energy relate to nonrelativistic, clamped-nuclei values with inner shell electrons constrained to be doubly occupied (C$_2$--C$_5$ from W4 theory, C$_6$ and C$_7$ from W3.2 theory, C$_8$ and C$_9$ from W3.2lite theory, C$_{10}$ from W2.2 theory).\\
$^b$ Highest doubly occupied molecular orbital\\
$^c$ HOMO-1: 1.904($\times$2), LUMO+1: 0.075($\times$2)\\
$^d$ cc-pVDZ basis set
\end{flushleft}
\end{table}

\clearpage

\squeezetable
\begin{table}
\caption{Component breakdown of the final W2.2 and W3.2lite total atomization energies at the bottom of the well (in kcal/mol) obtained from B3LYP/cc-pVTZ reference geometries\label{tab:W3.2L}}
\begin{tabular}{l|cccccccccc|cc}
\hline\hline
 & SCF & valence & valence & $\hat{T}_3-$(T) & 1.1$\times$(Q) & inner & relativ. & spin-orbit & DBOC & (a) & \multicolumn{2}{c}{TAE$_e$} \\
 & & CCSD & (T) & & /cc-pVDZ & shell &  &  &  & & W2.2 & W3.2lite \\
\hline
C$_2$($^1\Sigma_g^+$)   &  18.30  & 107.60 & 19.39 & -2.25 & 2.95 & 1.01 & -0.17 & -0.17 & 0.03  & 0.04 & 145.99   & 146.70  \\
C$_2$($^3\Pi_u$)        &  73.36  & 60.22  & 9.19  & -0.17 & 1.08 & 0.96 & -0.09 & -0.17 & 0.01  & 0.03 & 143.48   & 144.41  \\
C$_2$($^3\Sigma_g^-$)   &  86.79  & 36.07  & 4.27  &  0.39 & 0.32 & 0.67 & -0.07 & -0.17 & 0.00  & 0.03 & 127.56   & 128.28  \\
C$_3$($^1\Sigma_g^+$)   &  207.96 & 93.01  & 17.24 & -0.96 & 1.65 & 2.07 & -0.19 & -0.25 & 0.02  & 0.06 & 319.87   & 320.59  \\
C$_4$($^3\Sigma_g^-$)   &  288.48 & 125.46 & 19.76 & -1.14 & 1.99 & 3.21 & -0.33 & -0.34 & -0.29 & 0.08 & 435.95   & 436.84 \\
C$_4$($^1A_g$)          &  281.76 & 133.58 & 20.08 & -1.74 & 1.98 & 2.41 & -0.29 & -0.34 & 0.03  & 0.08 & 437.23   & 437.51  \\
C$_5$($^1\Sigma_g^+$)   &  404.38 & 167.41 & 31.90 & -2.51 & 4.01 & 4.62 & -0.47 & -0.42 & 0.07  & 0.10 & 607.50   & 609.05  \\
C$_6$($^3\Sigma_g^-$)   &  485.83 & 204.97 & 34.30 & -2.86 & 4.20 & 5.86 & -0.62 & -0.51 & -0.17 & 0.12 & 729.65   & 731.06 \\
C$_6$($^1A_1^\prime$)   &  476.56 & 227.68 & 36.67 & -3.32 & 4.13 & 6.84 & -0.87 & -0.51 & 0.06  & 0.13 & 746.43   & 747.29  \\
C$_7$($^1\Sigma_g^+$)   &  599.93 & 242.14 & 46.43 & -4.04 & 6.79 & 7.22 & -0.76 & -0.59 & 0.11  & 0.19 & 894.49   & 897.34  \\
C$_8$($^3\Sigma_g^-$)   &  683.35 & 282.36 & 48.82 & -4.79 & 6.80 & 8.50 & -0.92 & -0.68 & -0.10 & 0.16 & 1021.33  & 1023.43  \\
C$_8$($^1A_g$)          &  681.28 & 298.24 & 45.97 & -2.62 & 7.06 & 9.53 & -1.16 & -0.68 & 0.09  & 0.17 & 1033.28  & 1037.80  \\
C$_9$($^1\Sigma_g^+$)   &  795.86 & 317.23 & 61.02 & -6.44 & 9.79 & 9.83 & -1.06 & -0.76 & 0.15  & 0.19 & 1182.27  & 1185.72 \\
C$_{10}$($^3\Sigma_g^-$)&  880.78 & 359.06 & 63.43 & N/A  & N/A  & 11.13 & -1.21 & -0.85 & -0.02 & 0.21 & 1312.32  & N/A\\
C$_{10}$($^1A_1^\prime$)&  934.19 & 376.34 & 64.38 & N/A  & N/A  & 12.60 & -1.46 & -0.85 & -0.01 & 0.21 & 1385.20  & N/A\\
\hline
& \multicolumn{10}{c}{linear-cyclic isomerization energies} &&\\
\hline
C$_4$($^1A_g$)$\rightarrow$C$_4$($^3\Sigma_g^-$)       & -6.72 & 8.12  & 0.32  & -0.60 & -0.01 & -0.80 &  0.04 &0& 0.32 & 0.01 & 1.29 & 0.67\\
C$_6$($^1A_1^\prime$)$\rightarrow$C$_6$($^3\Sigma_g^-$)& -9.27 & 22.71 & 2.37  & -0.46 & -0.07 &  0.98 & -0.25 &0& 0.23 & 0.00 & 16.77 & 16.24\\
C$_8$($^1A_g$)$\rightarrow$C$_8$($^3\Sigma_g^-$)       & -2.07 & 15.88 & -2.86 & 2.17  & 0.26  &  1.03 & -0.24 &0& 0.19 & 0.00 & 11.94 & 14.37\\
C$_{10}$($^1A_1^\prime$)$\rightarrow$C$_{10}$($^3\Sigma_g^-$)& 53.41 & 17.28 & 0.95 & N/A & N/A & 1.48 & -0.25 & 0 & 0.01 & N/A & 72.88 & N/A\\
\hline
& \multicolumn{10}{c}{isodesmic reaction energies} &&\\
\hline
2C$_5\rightarrow$C$_7$+C$_3$     & 0.86 & -0.32 & 0.12 & -0.03 & -0.42 & -0.05 & 0.01 & 0.00 & 0.01 & -0.05 &0.62& 0.17\\
2C$_6\rightarrow$C$_8$+C$_4$     & -0.17 & 2.11 & 0.02 & 0.22 & -0.39 & 0.01 & 0.00 & 0.00 & 0.05 & 0.00 &2.02& 1.85\\
2C$_7\rightarrow$C$_9$+C$_5$     & -0.37 & -0.36 & -0.05 & 0.87 & -0.22 & 0.00 & 0.00 & 0.00 & 0.01 & 0.09 &-0.73&-0.08\\
2C$_8\rightarrow$C$_{10}$+C$_6$  & 	0.09 & 0.70 & -0.09 & N/A & N/A & 0.01 & 0.00 & 0.00 & -0.01 & 0.00 &0.70& N/A \\
3C$_5\rightarrow$C$_{9}$+2C$_3$ & 1.35 & -1.01 & 0.20 & 0.82 & -1.05 & -0.11 & 0.02 & 0.00 & 0.02 & 0.00 &0.48& 0.25\\
\hline\hline
\end{tabular}
\begin{flushleft}
(a) difference between the ACES II and MOLPRO definitions of the valence ROCCSD(T)\\
(b) RCCSDTQ/cc-pVDZ$-$UCCSDTQ/cc-pVDZ\\
\end{flushleft}
\end{table}
\clearpage

\squeezetable
\begin{table}
\caption{Component breakdown of the final W4$^c$ total atomization energies at the bottom of the well (in kcal/mol) obtained from CCSD(T)/cc-pVQZ reference geometries\label{tab:W4}}
\begin{tabular}{l|cccccccccccccc}
\hline\hline
 & SCF & valence & valence & $\hat{T}_3-$(T) & $\hat{T}_4$ & $\hat{T}_5$ & $\hat{T}_6$ & inner & relativ. & spin-orbit & DBOC & (a) & (b) & TAE$_e$$^a$ \\
 & & CCSD & (T) & & (c) & & & shell &  &  &  & &  \\
\hline
C$_2$($^1\Sigma_g^+$)   & 18.38  & 107.60 & 19.37 & -2.19 & 2.37 & 0.24    & [0] & 1.06 & -0.17 & -0.17 & 0.03  & 0.04 & 0    & 146.52\\
ditto W4.3              & 18.38  & 107.60 & 19.37 & -2.24 & 2.35 & 0.32    & 0.07 & 1.25 & -0.17 & -0.17 & 0.03  & 0.00 & 0   & 146.78 \\
C$_2$($^3\Pi_u$)        & 72.87  & 60.65  & 9.39  & -0.25 & 1.33 & 0.14    & [0] & 0.89 & -0.09 & -0.17 & 0.01  & 0.03 & 0.05 & 144.78\\
ditto W4.3              & 72.87  & 60.65  & 9.39  & -0.27 & 1.29 & 0.12    & 0.01 & 0.92 & -0.09 & -0.17 & 0.01  & 0.00 & 0.05   & 144.72 \\
C$_2$($^3\Sigma_g^-$)   & 86.62  & 36.29  & 4.35  &  0.25 & 0.41 & 0.03    & [0]  & 0.64 & -0.07 & -0.17 & 0.00 & 0.03 & 0.00 & 128.36 \\
ditto W4.3              & 86.62 & 36.29 & 4.35 & 0.22 & 0.42 & 0.03 & 0.001 & 0.65 & -0.07 & -0.17 & 0.00 & 0.00 & 0.00 & 128.34 \\
C$_3$($^1\Sigma_g^+$)   & 207.18 & 93.58  & 17.63 & -1.10 & 2.07 & -0.02   & [0] & 1.96 & -0.19 & -0.25 & 0.02  & 0.06 & 0    & 320.90\\
C$_4$($^3\Sigma_g^-$)   & 287.62 & 126.19 & 20.06 & -1.37 & 2.56 & 0.12    & [0] & 3.08 & -0.34 & -0.34 & -0.35 & 0.08 & 0.33 & 437.28\\
C$_4$($^1A_g$)          & 280.97 & 134.15 & 20.48 & -1.91 & 2.16 & 0.19    & [0] & 2.28 & -0.29 & -0.34 & 0.03  & 0.08 & 0    & 437.77\\
C$_5$($^1\Sigma_g^+$)   & 403.09 & 168.48 & 32.47 & -2.75 & 3.39 & 0.21$^b$& [0] & 4.42 & -0.48 & -0.42 & 0.07  & 0.10 & 0    & 608.54\\
C$_6$($^3\Sigma_g^-$)$^c$ & 484.46 & 206.00 & 34.80 & -3.19 & 3.92 & N/A &  [0] &5.71 & -0.64 & -0.51 & -0.23 & 0.12 & N/A & 715.89\\
C$_6$($^1A_1^\prime$)$^c$ & 475.87 & 228.16 & 37.07 & -3.60 & 3.85 & N/A &  [0] &6.58 & -0.88 & -0.51 &  0.06 & 0.13 & N/A & 729.68\\
C$_7$($^1\Sigma_g^+$)$^c$ & 589.19 & 243.37 & 47.20 & -4.68 & 6.37 & N/A &  [0] &6.96 & -0.78 & -0.59 &  0.12 & 0.15 & 0   & 896.22\\
\hline
& \multicolumn{13}{c}{singlet-triplet and linear-cyclic energy differences}\\
\hline
C$_2$($X~^1\Sigma^+_g$)$\rightarrow$C$_2$($a~^3\Pi_u$) & -54.49 & 46.95 & 9.98 & -1.94 & 1.04 & 0.10 & [0] & 0.17 & -0.08 & 0 & 0.01 & 0.01 & -0.05   & 1.74\\
ditto W4.3 & -54.49 & 46.95 & 9.98 & -1.96 & 1.06 & 0.20 & 0.06 & 0.33 & -0.08 & 0 & 0.01 & 0 & -0.05   & 2.05$^d$\\
C$_2$($X~^1\Sigma^+_g$)$\rightarrow$C$_2$($^3\Sigma_g^-$) & -68.24 & 71.31 & 15.02 & -2.44 & 1.96 & 0.21 & [0] & 0.41 & -0.10 & 0 & 0.02 & 0.01 & 0.00   &18.16 \\
ditto W4.3 & -68.24 & 71.31 & 15.02 & -2.46 & 1.93 & 0.29 & 0.07 & 0.59 & -0.10 & 0 & 0.02 & 0 & 0.00   & 18.44$^e$\\
C$_4$($^1A_g$)$\rightarrow$C$_4$($^3\Sigma_g^-$) & -6.64 & 7.96 & 0.41 & -0.54 & -0.40 & 0.07 &  [0] & -0.80 & 0.05 & 0 & 0.38 & 0.01 & -0.33 & 0.49\\
C$_6$($^1A_1^\prime$)$\rightarrow$C$_6$($^3\Sigma_g^-$)& -8.60 & 22.16 & 2.27 & -0.41 & -0.08 & N/A &  [0] & 0.87 & 0.24 & 0 & 0.29 & 0.00 & N/A & 16.26\\
\hline\hline
\end{tabular}
\begin{flushleft}
(a) difference between the ACES II and MOLPRO definitions of the valence ROCCSD(T)\\
(b) RCCSDTQ/cc-pVDZ$-$UCCSDTQ/cc-pVDZ\\
(c) UHF reference: for RCCSDTQ$-$UCCSDTQ, cfr. (b)\\
$^a$ Note that the TAE$_e$ do not include $\Delta$DBOC\\
$^b$$\hat{T}_5$ approx. as CCSDTQ(5)$_\Lambda$/cc-pVDZ(no d)$-$CCSDTQ/cc-pVDZ(no d).\\
$^c$C$_6$ and C$_7$ from W3.2 theory.\\
$^d$Experimental value: 2.05 kcal/mol\cite{Hub79}.\\
$^e$Experimental value: 18.40 kcal/mol\cite{Hub79}.\\
\end{flushleft}
\end{table}
\clearpage

\squeezetable
\begin{table}
\caption{Total atomization energies at 0 K (kcal/mol)$^a$\label{tab:TAE0}}
\begin{tabular}{ll|c|cccc|c|cccccc}
\hline\hline 
& & ZPVE$^b$ & W2.2 & W3.2lite & W3.2 & W4lite & W4  &\multicolumn{2}{c}{MT}& MFG & RB & Exp.$^c$ & uncert. \\ 
&&&&&&&&cc-pVTZ&cc-pVQZ&&&&\\
\hline
C$_2$($^1\Sigma_g^+$)   &    & 2.64$^d$  & 143.39  & 144.06  & 143.88 & 143.93 & 143.88$^e$ & 145.1  & 144.2 & 144.8 & 147.1 & 144.6 &1.9\\
C$_2$($^3\Pi_u$)        &    & 2.34$^d$  & 141.16  & 142.07  & 141.94 & 141.99 & 142.45$^f$ & 141.1  & 142.4 &       &       &           &\\
C$_2$($^3\Sigma_g^-$)   &    & 2.09$^d$  & 125.45 & 126.19 & 126.01 & 126.13 & 126.27\\
C$_3$($^1\Sigma_g^+$)   & lin& 5.07  & 314.80  & 315.52  & 315.27 & 315.32 & 315.83 & 314.4  & 314.3 & 315.7 & 320.9 & 311.4     &3.1\\
C$_4$($^3\Sigma_g^-$)   & lin& 8.11  & 427.83  & 428.72  & 428.26 & 428.34 & 429.16 & 423.5  & 426.1 &       &       & 428.5     &4.1\\
C$_4$($^1A_g$)          & cyc& 7.68  & 429.55  & 429.83  & 429.52 & 429.60 & 430.09 & 427.7  & 429.3 & 430.3 & 437.6 &           &\\
C$_5$($^1\Sigma_g^+$)   & lin& 11.89 & 595.60  & 597.16  & 596.65 & 596.81 & 596.64 & 592.9  & 594.0 & 596.8 & 606.8 & 591.5     &4.1\\
C$_6$($^3\Sigma_g^-$)   & lin& 14.50 & 715.15  & 716.56  & 715.89 &716.01$^g$&717.19$^g$& 708.0  & N/A   &       &       & 706.3     &4.8\\
C$_6$($^1A_1^\prime$)   & cyc& 16.98 & 729.45  & 730.31  & 729.68 &        &        & 720.6  & 724.9 & 722.8 & 734.7 &           &\\
C$_7$($^1\Sigma_g^+$)   & lin& 18.18 & 876.31  & 879.16[878.8]  & 878.03& 878.28$^h$& 877.45$^h$& 870.9  &[873.7]& 877.7 & 892.4 & 873.1     &4.8\\
C$_8$($^3\Sigma_g^-$)   & lin& 20.70 & 1000.63 & 1002.73[1004.4] & 1001.86$^i$&        &        & 992.9  & N/A   &       &       & [984.0]   &[10.4]\\
C$_8$($^1A_g$)          & cyc& 22.83 & 1010.44 & 1014.97 &        &        &        & 998.9  & N/A   & 1004.6& 1021.3&           &\\
C$_9$($^1\Sigma_g^+$)   & lin& 24.35 & 1157.91 & 1161.37[1161.2] & 1160.36$^j$ & 1160.72$^j$ & 1159.21$^j$ & 1149.6 & N/A   & 1158.4& 1177.8& [1154.6]  &[10.4]\\
C$_{10}$($^3\Sigma_g^-$)& lin& 26.82 & 1285.50 & 1288.22$^k$[1288.9]&        &        &        &[1277.8]& N/A   &       &       &           &\\
C$_{10}$($^1A_1^\prime$)& cyc& 29.65 & 1355.54 &         &        &        &        & 1343.1 & N/A   & 1344.3& 1366.6&           &\\
\hline
\multicolumn{14}{c}{Reaction energies (linear clusters)}\\
\hline
\multicolumn{3}{c}{2C$_4\rightarrow$C$_6$+C$_2$($^3\Pi_u$)}     & -0.80  & -1.18 & -1.32 &  &  & -2.10  &  & & &  &\\
\multicolumn{3}{c}{2C$_4\rightarrow$C$_6$+C$_2$($^3\Sigma_g^-$)}& 14.91 & 14.70 & 14.62 &  &  &   &  & & &  &\\
\multicolumn{3}{c}{2C$_5\rightarrow$C$_7$+C$_3$}     & 0.09  & -0.37 & 0.01 &  &  & 0.5  &  & & & -1.43 &9.9\\
\multicolumn{3}{c}{2C$_6\rightarrow$C$_8$+C$_4$}     & 1.84  & 1.67  &  &  &  & -0.4 &  & & &       &\\
\multicolumn{3}{c}{2C$_7\rightarrow$C$_9$+C$_5$}     & -0.94 & -0.20 &  &  &  & -0.7 &  & & &       &\\
\multicolumn{3}{c}{2C$_8\rightarrow$C$_{10}$+C$_6$}  & 0.67  &       &  &  &  &      &  & & &       &\\
\multicolumn{3}{c}{3C$_5\rightarrow$C$_{9}$+2C$_3$}  & -0.75 & -0.94 &  &  &  &      &  & & &       &\\
\hline\hline
\end{tabular}
\begin{flushleft}
$^a$Reference geometries for all the W3.2lite calculations and for all clusters larger than C$_7$ are at the B3LYP/cc-pVTZ level of theory; values in square parenthesis were obtained with the formula: TAE$_0$[C$_{n}$]$=$2$\times$TAE$_0$[C$_{n-2}$]$-$TAE$_0$[C$_{n-4}$].\\
$^b$The zero-point vibrational energies for C$_3-$C$_{10}$ are B3LYP/pc-2 scaled by 0.985.\\
$^c$Ref. \cite{Ging94}.\\
$^d$From $\omega_e/2-\omega_ex_e/4$, with $\omega_e$ and $\omega_ex_e$ taken from Ref.\cite{Hub79}.\\
$^e$ATcT value is 144.03$\pm$0.13 kcal/mol, W4.2 value is 144.05 kcal/mol (142.46 kcal/mol for the triplet state).\\
W4.3 and W4.4 values for the singlet state are 144.08 and 144.07 kcal/mol, respectively\cite{W4.4}.\\
$^f$W4.3 value is 142.39 kcal/mol.\\
$^g$Estimated by assuming that the 2C$_4$($^3\Sigma_g^-$)$\rightarrow$C$_6$($^3\Sigma_g^-$)+C$_2$($^3\Pi_u$) reaction energy remains unchanged at the W3.2 and post-W3.2 levels. Via C$_2$($^3\Sigma_g^-$), the estimated W4lite and W4 numbers are 715.93 and 717.43 kcal/mol, respectively.\\
$^h$Estimated by assuming that the 2C$_5\rightarrow$C$_7$+C$_3$ reaction energy remains unchanged at the W3.2 and post-W3.2 levels.\\
$^i$Estimated by assuming that the 2C$_6\rightarrow$C$_8$+C$_4$ reaction energy remains unchanged at the W3.2lite and W3.2 levels.\\
$^j$Estimated by assuming that the 3C$_5\rightarrow$C$_9$+2C$_3$ reaction energy remains unchanged at the W3.2lite and post-W3.2lite levels.\\
$^k$Estimated by assuming that the 2C$_8\rightarrow$C$_{10}$+C$_6$ reaction energy remains unchanged at the W2.2 and W3.2lite levels.\\
\end{flushleft}
\end{table}
\clearpage

\squeezetable
\begin{table}
\caption{Performance of standard composite computational thermochemistry methods. Total atomization energies at 0 K and error statistics with respect to our best values (in kcal/mol)\label{tab:Gn}}
\begin{tabular}{l|cccccccccc|c}
\hline\hline 
& G2MP2  & G2     & G3      & G3B3    & CBS-QB3 & CBS-APNO & G4  & G4(MP2) &     W1h    & W2.2 & Best estimate\cite{DFT_ref_val}\\
\hline
C$_2$($^1\Sigma_g^+$)   & 146.81 & 146.90 & 146.94  & 147.91  & 144.29  & 144.28   & 147.01  & 148.89 & 143.15  & 143.39 & 144.07\\
C$_2$($^3\Pi_u$)        & 140.28 & 140.48 & 142.72  & 143.73  & 140.13  & 140.83   & 142.35  & 143.46 & 141.00  & 141.16 & 142.39\\
C$_3$($^1\Sigma_g^+$)   & 316.75 & 316.89 & 317.53  & 318.86  & 314.21  & 314.96   & 316.93  & 318.84 & 314.75  & 314.79 & 315.83\\
C$_4$($^1A_g$)          & 430.30 & 430.73 & 428.98  & 430.91  & 427.39  & 430.12   & 431.66  & 433.94 & 429.10  & 429.52 & 430.09\\
C$_4$($^3\Sigma_g^-$)   & 424.52 & 425.63 & 429.61  & 431.65  & 426.87  & 428.10   & 428.85  & 429.67 & 428.15  & 427.73 & 429.16\\
C$_5$($^1\Sigma_g^+$)   & 596.95 & 597.50 & 599.98  & 601.87  & 594.52  & 597.19   & 598.34  & 600.44 & 595.71  & 595.58 & 596.64\\
C$_6$($^1A_1^\prime$)   & 725.51 & 726.01 & 727.73  & 732.43  & 727.18  & 731.10   & 731.72  & 733.92 & 728.76  & 729.36 & 729.68\\
C$_6$($^3\Sigma_g^-$)   & 710.69 & 709.14 & 718.39  & 719.34  & 714.86  & 719.04   & 715.19  & 715.87 & 715.44  & 715.10 & 717.19\\
C$_7$($^1\Sigma_g^+$)   & 876.39 & 877.42 & 881.72  & 884.02  & 874.30  & 878.83   & 879.28  & 881.41 & 876.45  & 876.28 & 877.45\\
C$_8$($^1A_g$)          & 1005.15& 1005.89& 1011.16 & 1016.63 & 1006.61 &          & 1013.39 & 1015.80 & 1009.80 & 1010.44 & 1014.97\\
C$_8$($^3\Sigma_g^-$)   & 994.13 & 996.90 & 1004.38 & 1004.68 & 1000.48 &          & 999.09  & 999.66 & 1000.90 & 1000.63 & 1001.86\\
C$_9$($^1\Sigma_g^+$)   & 1156.44& 1158.01& 1164.12 & 1166.88 & 1155.66 &          & 1160.51 & 1162.64 & 1158.07 & 1157.91 & 1159.21\\
C$_{10}$($^1A_1^\prime$)& 1350.14& 1351.63& 1359.15 & 1363.49 & 1353.80 &          & 1359.14 & 1361.25 & 1355.11 & 1355.54 & 1355.54\\
C$_{10}$($^3\Sigma_g^-$)& 1277.31& 1280.91& 1290.12 & 1289.43 & 1284.83 &          & 1283.48 & 1283.95 & 1285.72 & 1285.50 & 1288.22\\
RMSD    &  5.39 & 4.49 & 2.79 & 4.22 & 3.22 &      & 2.30 & 3.45 & 1.83 & 1.76\\
RMSD$^a$&  4.65 & 4.22 & 2.77 & 3.92 & 3.30 &      & 1.80 & 3.11 & 1.84 & 1.73\\
RMSD$^b$&  3.25 & 3.41 & 2.29 & 3.58 & 2.27 & 1.16 & 1.73 & 3.31 & 1.14 & 1.17\\
RMSD$^c$&  3.31 & 3.48 & 2.21 & 3.55 & 2.41 & 1.22 & 1.51 & 3.07 & 1.17 & 1.22\\
\hline\hline
\end{tabular}
\begin{flushleft}
$^a$RMSD w/o C$_{10}$.\\
$^b$RMSD w/o C$_8$--C$_{10}$.\\
$^c$RMSD w/o C$_8$--C$_{10}$ and C$_2$($^1\Sigma_g^+$).\\
\end{flushleft}
\end{table}
\clearpage

\clearpage
\squeezetable
\begin{table}
\caption{Performance statistics (kcal/mol) of various exchange-correlation functionals for the carbon clusters considered in the present work.$^a$ Unless otherwise indicated all calculations were done with the pc-2 basis set.\label{tab:DFT}}
\resizebox{0.40\textwidth}{!}{%
\begin{tabular}{l|l|cccccc}
\hline\hline
 Class & Functional & RMSD & MSD & MAD & RMSD$^a$ & RMSD$^b$\\
\hline
   & HCTH407  &32.7 & 25.9 & 27.8 & 33.7 & 15.3\\
   & BLYP     &29.2 & 22.0 & 23.9 & 30.1 & 10.5\\
GGA& BPW91    &48.5 & 40.5 & 41.8 & 50.2 & 26.8\\
   & BP86     &60.4 & 51.6 & 52.3 & 62.7 & 35.7\\
   & PBE      &74.4 & 63.9 & 64.3 & 77.2 & 46.8\\
\hline
    & M06-L       &43.3 & 35.6 & 37.4 & 44.8 & 23.5\\
meta& VSXC        &31.8 & 24.8 & 26.7 & 32.8 & 13.0\\
GGA & TPSS        &22.6 & 15.6 & 18.2 & 22.9 & 7.3\\
    & $\tau$HCTH  &27.9 & 21.2 & 23.4 & 28.6 & 11.2\\
\hline
      & PBE0   &28.1 & 20.1 & 23.9 & 28.3 & 14.1\\
      & B97-2  &25.5 & 18.6 & 21.8 & 25.7 & 11.2\\
      & B3PW91 &21.9 & 14.3 & 18.0 & 21.6 & 8.1\\
hybrid& B97-1  &22.1 & 14.7 & 18.3 & 21.8 & 7.8\\
GGA   & B98    &14.5 & 5.3 & 11.7 & 12.8 & 5.5\\
      & TPSSh  &12.2 & 1.7 & 9.7 & 10.3 & 7.1\\
      & B3LYP  &13.1 & 2.0 & 10.9 & 11.3 & 8.2\\
      & mPW1K  &20.2 & -17.7 & 17.7 & 17.0 & 17.2\\
      & BHLYP  &56.6 & -52.4 & 52.4 & 56.6 & 48.1\\
\hline
      & mPW1B95    &20.7 & 13.6 & 17.7 & 20.3 & 9.4\\
      & B1B95      &18.1 & 11.2 & 15.1 & 17.4 & 7.3\\
      & M06        &19.7 & 14.2 & 17.2 & 19.6 & 11.1\\
hybrid& PW6B95     &18.1 & 10.9 & 14.9 & 17.4 & 6.3\\
meta  & TPSS1KCIS  &23.2 & 16.2 & 19.2 & 23.3 & 8.3\\
GGA   & M05        &10.4 & 4.3 & 8.7 & 9.4 & 8.8\\
      & $\tau$HCTHh&24.1 & 16.8 & 20.1 & 24.2 & 9.1\\
      & BMK        &11.8 & -4.2 & 7.5 & 6.2 & 7.6\\
      & M06-2X     &10.5 & 2.5 & 7.7 & 7.5 & 4.2\\
      & BB1K       &12.7 & -8.9 & 9.0 & 8.4 & 9.9\\
      & PWB6K      &14.6 & -11.1 & 11.1 & 10.4 & 11.7\\
\hline
       & B2GP-PLYP$^c$ & 9.1 & 3.4 & 7.0 & 8.6 & 2.6\\
       & B2GP-PLYP$^d$ & 12.1 & 7.1 & 9.3 & 12.1 & 2.1\\
       & B2GP-PLYP$^e$ & 7.7 & 1.5 & 6.0 & 6.9 & 3.8\\
       & B2GP-PLYP$^f$ & 10.2 & 5.1 & 7.8 & 9.9 & 1.8\\
       & B2-PLYP$^c$   & 13.8 & 8.7 & 10.7 & 14.0 & 2.6\\
       & B2-PLYP$^d$   & 16.4 & 11.4 & 12.9 & 16.7 & 4.1\\
       & B2-PLYP$^e$   & 12.2 & 7.1 & 9.6 & 12.2 & 2.4\\
       & B2-PLYP$^f$   & 14.6 & 9.7 & 11.4 & 14.8 & 3.1\\
       & B2T-PLYP$^c$  & 9.2 & 3.3 & 7.4 & 8.6 & 3.2\\
double & B2T-PLYP$^d$  & 11.7 & 6.4 & 9.1 & 11.5 & 2.0\\
hybrid & B2T-PLYP$^e$  & 8.0 & 1.5 & 6.5 & 7.2 & 4.3\\
       & B2T-PLYP$^f$  & 10.0 & 4.6 & 7.9 & 9.6 & 2.4\\
       & B2K-PLYP$^c$  & 7.4 & 0.4 & 5.7 & 6.4 & 4.0\\
       & B2K-PLYP$^d$  & 10.1 & 4.8 & 7.5 & 9.8 & 1.9\\
       & B2K-PLYP$^e$  & 6.7 & -1.8 & 5.2 & 5.5 & 5.5\\
       & B2K-PLYP$^f$  & 8.2 & 2.5 & 6.2 & 7.5 & 2.5\\
       & mPW2-PLYP$^c$ & 8.1 & -0.1 & 6.7 & 6.9 & 6.2\\
       & mPW2-PLYP$^d$ & 9.2 & 2.4 & 7.7 & 8.3 & 4.4\\
       & mPW2-PLYP$^e$ & 7.8 & -1.6 & 5.9 & 6.3 & 7.2\\
       & mPW2-PLYP$^f$ & 8.3 & 0.8 & 6.9 & 7.2 & 5.4\\
\hline\hline
\end{tabular}}
\begin{flushleft}
$^a$Excluding C$_2$($^1\Sigma_g^+$).\\
$^b$Excluding C$_2$($^1\Sigma_g^+$), C$_{10}$, and all the linear systems larger than C$_4$.\\
$^c$(all electron) cc-pwCVQZ basis set.\\
$^d$(all electron) cc-pwCVQZ basis set combined with a CBS extrapolation where Nmin=15 as recommended in Ref. \cite{Petersson}.\\
$^e$(frozen core) pc-3 basis set.\\
$^f$(frozen core) pc-3 basis set combined with a CBS extrapolation where Nmin=15 as recommended in Ref. \cite{Petersson}.\\
\end{flushleft}
\end{table}
\clearpage

\clearpage


\begin{thebibliography}{99}

\bibitem{Say98} Van Orden, A.; Saykally, R. J. {\em Chem. Rev.} {\bf 98}, 2313 (1998) 

\bibitem{Lif00} Lifshitz, C. {\em Int. J. Mass Spectrom.} {\bf 200}, 423 (2000)

\bibitem{Ohn08}  
S. Arulmozhiraja and T. Ohno,
{\em J. Chem. Phys.} {\bf 128}, 114301 (2008)

\bibitem{Rag87} 
(a) Raghavachari, K.; Binkley, J. S. {\em J. Chem. Phys.} {\bf 1987}, {\it 87}, 2191;
(b) Raghavachari, K. Personal communication

\bibitem{HLD} Hutter, J.; L\"uthi, H. P.; Diederich, F. {\em J. Am. Chem. Soc.} {\bf 116}, 750 (1994)

\bibitem{Mar96} Martin, J. M. L.; Taylor, P. R. {\em J. Phys. Chem. A} {\bf 100}, 6047 (1996)



\bibitem{exptC6} 
(a) Presilla-Marquez, J. D.; Sheehy, J. A.; Mills, J. D.; Carrick, P. G.; Larson, C. W. {\em Chem. Phys. Lett.} {\bf 274}, 439 (1997); 
(b) Wang, S. L.; Rittby, C. M. L.; Graham, W. R. M. {\em J. Chem. Phys.} {\bf 107}, 6032 (1997)

\bibitem{exptC8} 
(a) Wang, S. L.; Rittby, C. M. L.; Graham, W. R. M. {\em J. Chem. Phys.} {\bf 107}, 7025 (1997); 
(b) Presilla-Marquez, J. D.; Harper, J.; Sheehy, J. A.; Carrick, P. G.; Larson, C. W. {\em Chem. Phys. Lett.} {\bf 300}, 719 (1999) 

\bibitem{Zaj92} Zajfman, D.; Feldman, H.; Heber, O.; Kella, D.; Majer, D.; Vager, Z.;  Naaman, R. {\em Science} {\bf 258}, 1129 (1992) 

\bibitem{C4C5expt} 
(a) Vala, M.; Chandrasekhar, T. M.; Szczepanski, J.; Van Zee, R. J.; Weltner, W., Jr. {\em J. Chem. Phys.} {\bf 90}, 595 (1989);
(b) Bernath, P. F.; Hinkle, K. H.; Keady, J. J. {\em Science} {\bf 244}, 562 (1989); Heath, J. R.; Cooksy, A. L.; Gruebele, M. H. W.; Schmuttenmaer, C. A.; Saykally, R. J. {\em Science} {\bf 244}, 564 (1989); Moazzen-Ahmadi, N.; McKellar, A. R. W.; Amano, T. {\em J. Chem. Phys.} {\bf 91}, 2140 (1989); Moazzen-Ahmadi, N.; McKellar, A. R. W.;  Amano, T. {\em Chem. Phys. Lett.} {\bf 157}, 1 (1989); 
(c) Shen, L. N.; Graham, W. R. M. {\em J. Chem. Phys.} {\bf 91}, 5115 (1989); Heath, J. R.; Saykally, R. J. {\em J. Chem. Phys.} {\bf 94}, 3271 (1991)

\bibitem{C4C5calcs} 
(a) Michalska, D.; Chojnacki, H.; Hess, B. A., Jr.; Schaad, L. {\em J. Chem. Phys. Lett.} {\bf 141}, 376 (1987);
(b) Bernholdt, D. E.; Magers, D. H.; Bartlett, R. J. {\em J. Chem. Phys.} {\bf 89}, 3612 (1988); 
(c) Martin, J. M. L.; Francois, J. P.; Gijbels, R. {\em J. Chem. Phys.} {\bf 90}, 3403 (1989)
 
\bibitem{C7C9calc}
(a) Martin, J. M. L.; Francois, J. P.; Gijbels, R. {\em J. Chem. Phys.} {\bf 93}, 8850 (1990); 
(b) Martin, J. M. L.; Francois, J. P.; Gijbels, R. {\em J. Comput. Chem.} {\bf 12}, 52 (1991) 

\bibitem{C6expt}
Vala, M.; Chandrasekhar, T. M.; Szczepanski, J.; Pellow, R. In Materials chemistry at high temperatures; J., Hastie, Ed.; Humana Press: 
Clifton, NJ, 1990; Vol II. Kranze, R. H.; Graham, W. R. M. J. Chem. 
Phys. 1993, 98, 71. Hwang, H. J.; Vanorden, A.; Tanaka, K.; Kuo, E. W.; 
Heath, J. R.; Saykally, R. J. Mol. Phys. 1993, 79, 769. 
Kranze, R. H.; Withey, P. A.; Rittby, C. M. L.; Graham, W. 
R. M. J. Chem. Phys., in press (1996).

\bibitem{C7C9expt}
(a) Heath, J. R.; Sheeks, R. A.; Cooksy, A. L.; Saykally, R. J. {\em Science} {\bf 249}, 895 (1990); Heath, J. R.; Vanorden, A.; Kuo, E.; Saykally, R. J. {\em Chem. Phys. Lett.} {\bf 182}, 17 (1991); Heath, J. R.; Saykally, R. J. {\em J. Chem. Phys.} {\bf 94}, 1724 (1991); 
(b) Heath, J. R.; Saykally, R. J. {\em J. Chem. Phys.} {\bf 93}, 8392 (1990); Vanorden, A.; Hwang, H. J.; Kuo, E. W.; Saykally, R. J. {\em J. Chem. Phys.} 98, 6678 (1993).

\bibitem{Mar96c4} Martin, J. M. L.; Schwenke, D. W.; Lee, T. J. and Taylor, P. R. {\em J. Chem. Phys.} {\bf 104}, 4657 (1996) and references therein

\bibitem{Schwarz2000} S. J. Blanksby, D. Schr\"oder, S. Dua, J. H. Bowie, and H. Schwarz,
{\em J. Am. Chem. Soc.} {\bf 122}, 7105 (2000).

\bibitem{Mar95c18} Martin, J. M. L.; El-Yazal, J. and Fran\c{c}ois, J.-P. {\em Chem. Phys. Lett.} {\bf 242}, 570 (1995)

\bibitem{Dro59} Drowart, J.; Burns, R. P.; DeMaria, G.; Inghram, M. G. {\em J. Chem. Phys.} {\bf 31}, 1131 (1959)

\bibitem{Mar91} Martin, J. M. L.; Fran\c{c}ois, J.-P.; Gijbels, R. {\em J. Chem. Phys.} {\bf 95}, 9420 (1991)

\bibitem{Ging94} Gingerich, K. A.; Finkbeiner, H. C.; Schmude, R. W., Jr. {\em J. Am. Chem. Soc.} {\bf 116}, 3884 (1994)

\bibitem{Mart95} Martin, J. M. L.; Taylor, P. R. {\em J. Chem. Phys.} {\bf 102}, 8270 (1995)

\bibitem{W4} Karton, A.; Rabinovich, E.; Martin, J. M. L.; Ruscic, B. {\em J. Chem. Phys.} {\bf 125}, 144108 (2006)

\bibitem{ATcT} 
(a) Ruscic, B. unpublished results obtained from Active Thermochemical Tables (ATcT) version 1.25 using the Core Argonne Thermochemical Network version 1.056 (2006);
(b) Ruscic, B.; Pinzon, R. E.; Morton, M. L.; Srinivasan, N. K.; Su, M.-C.; Sutherland, J. W.; Michael, J. V. {\em J. Phys. Chem. A} {\bf 110}, 6592 (2006);
(c) Tasi, G.; Izs\'ak, R.; Matisz, G.; Cs\'asz\'ar, A. G.; K\'allay, M.; Ruscic, B.; Stanton, J. F. {\em Chem. Phys. Chem.} {\bf 7}, 1664 (2006);
(d) Ruscic, B.; Pinzon, R. E.; Morton, M. L.; von Laszewski, G.; Bittner, S.; Nijsure, S. G.; Amin, K. A.; Minkoff, M.; Wagner, A. F. {\em J. Phys. Chem. A} {\bf 108}, 9979 (2004);
(e) Ruscic, B. ``Active Thermochemical Tables'', in: {\em 2005 Yearbook of Science and Technology} (annual update to {\em McGraw-Hill Encyclopedia of Science and Technology}), McGraw-Hill, New York (2004), pp 3-7

\bibitem{Liang90} C. X. Liang and H. F. Schaefer III, {\em Chem. Phys. Lett.} {\bf 169}, 150 (1990)

\bibitem{SchaeferC10} C. X. Liang and H. F. Schaefer III, {\em J. Chem. Phys.} {\bf 93}, 8844 (1990)

\bibitem{Grev90} R. S. Grev, I. L. Alberts, and H. F. Schaefer III, {\em J. Phys. Chem. A} {\bf 94}, 3379 (1990); {\em erratum} {\bf 94}, 8744 (1990)

\bibitem{SchaeferC2} M. L. Leininger, C. D. Sherrill, W. D. Allen, and H. F. Schaefer III, {\em J. Chem. Phys.} {\bf 108}, 6717 (1998)

\bibitem{Schaefer2007} L. Belau, S. E. Wheeler, B. W. Ticknor, M. Ahmed, S. R. Leone, W. D. Allen, H. F. Schaefer III, and M. A. Duncan, {\em J. Am. Chem. Soc.} {\bf 129}, 10229 (2007).


\bibitem{Rag89Wat93} 
(a) Raghavachari, K.; Trucks, G. W.; Pople, J. A.; Head-Gordon, M. {\em Chem. Phys. Lett.} {\bf 157}, 479 (1989); 
(b) Watts, J. D.; Gauss, J.; Bartlett, R. J. {\em J. Chem. Phys.} {\bf 98}, 8718 (1993)

\bibitem{molpro} MOLPRO is a package of ab initio programs written by H.-J. Werner, P. J. Knowles, M. Sch\"utz, R. Lindh, P. Celani, T. Korona, G. Rauhut, F. R. Manby, R. D. Amos, A. Bernhardsson, A. Berning, D. L. Cooper, M. J. O. Deegan, A. J. Dobbyn, F. Eckert, C. Hampel, G. Hetzer, A. W. Lloyd, S. J. McNicholas, W. Meyer, M. E. Mura, A. Nickla§, P. Palmieri, R. Pitzer, U. Schumann, H. Stoll, A. J. Stone R. Tarroni, and T. Thorsteinsson

\bibitem{mrcc} MRCC, a string-based general coupled cluster program suite written by M. K\'allay. See also M. K\'allay and P. R. Surj\'an, J. Chem. Phys. {\bf 115}, 2945 (2001) as well as: \url{http://www.mrcc.hu}

\bibitem{aces2de} ACES II (Austin-Mainz-Budapest version) is an electronic structure program system written by J.F. Stanton, J. Gauss, J.D. Watts, P.G. Szalay, and R.J. Bartlett, with contributions from A.A. Auer, D.B. Bernholdt, O. Christiansen, M.E. Harding, M. Heckert, O. Heun, C. Huber, D. Jonsson, J. Jus\'elius, W.J. Lauderdale, T. Metzroth, and K. Ruud

\bibitem{psi3} 
T. D. Crawford, C. D. Sherrill, E. F. Valeev, J. T. Fermann, R. A. King, M. L. Leininger, S. T. Brown,
C. L. Janssen, E. T. Seidl, J. P. Kenny, and W. D. Allen, {\em J. Comput. Chem.} {\bf 28} 1610 (2007).
See also: \url{http://www.psicode.org}



\bibitem{Dun89} Dunning, T. H. {\em J. Chem. Phys.} {\bf 90}, 1007 (1989)



\bibitem{pwCVnZ} Peterson, K. A.; Dunning, T. H. {\em J. Chem. Phys.} {\bf 117}, 10548 (2002)

\bibitem{Dyall} de Jong, W. A.; Harrison, R. J.; Dixon, D. A. {\em J. Chem. Phys.} {\bf 114}, 48 (2001)

\bibitem{Kar-Mar} Karton A.; Martin, J. M. L. {\em Theor. Chem. Acc.} {\bf 115}, 330 (2006)

\bibitem{Jen05} Jensen, F. {\em Theor. Chem. Acc.} {\bf 113}, 267 (2005)

\bibitem{Klop01} Klopper, W. {\em Mol. Phys.} {\bf 99}, 481 (2001)

\bibitem{W4.4} Karton, A.; Taylor, P. R.; Martin, J. M. L. {\em J. Chem. Phys.} {\bf 127}, 064104 (2007)

\bibitem{W2} J. M. L. Martin and G. de Oliveira, {\em J. Chem. Phys.} {\bf 111}, 1843 (1999); J. M. L. Martin and S. Parthiban, {\em J. Chem. Phys.} {\bf 114}, 6014 (2001).

\bibitem{W1h} J. M. L. Martin and S. Parthiban, in {\em Quantum-Mechanical Prediction of Thermochemical Data} (ed. J. Cioslowski),
{\em Understanding Chemical Reactivity}, Vol. 22 (Kluwer, Dordrecht, 2001). Available online at:
\url{http://dx.doi.org/10.1007/0-306-47632-0_2}

\bibitem{W3} Boese, A. D.; Oren, M.; Atasoylu, O.; Martin, J. M. L.; K\'allay, M.; Gauss, J. {\em J. Chem. Phys.} {\bf 120}, 4129 (2004)

\bibitem{W3.2lite} Karton, A.; Kaminker, I.; Martin, J. M. L. {\em J. Phys. Chem. A}, to be published (R. B. Gerber issue).



\bibitem{DKH1DKH2} 
(a) Douglas, M.; Kroll, N. M. {\em Ann. Phys.} (NY) {\bf 82}, 89 (1974); 
(b) Hess, B. A. {\em Phys. Rev. A} {\bf 33}, 3742 (1986)

\bibitem{g03} Gaussian 03, Revision C.01, M. J. Frisch, G. W. Trucks, H. B. Schlegel, G. E. Scuseria, M. A. Robb, J. R. Cheeseman, J. A. Montgomery, Jr., T. Vreven, K. N. Kudin, J. C. Burant, J. M. Millam, S. S. Iyengar, J. Tomasi, V. Barone, B. Mennucci, M. Cossi, G. Scalmani, N. Rega, G. A. Petersson, H. Nakatsuji, M. Hada, M. Ehara, K. Toyota, R. Fukuda, J. Hasegawa, M. Ishida, T. Nakajima, Y. Honda, O. Kitao, H. Nakai, M. Klene, X. Li, J. E. Knox, H. P. Hratchian, J. B. Cross, V. Bakken, C. Adamo, J. Jaramillo, R. Gomperts, R. E. Stratmann, O. Yazyev, A. J. Austin, R. Cammi, C. Pomelli, J. W. Ochterski, P. Y. Ayala, K. Morokuma, G. A. Voth, P. Salvador, J. J. Dannenberg, V. G. Zakrzewski, S. Dapprich, A. D. Daniels, M. C. Strain, O. Farkas, D. K. Malick, A. D. Rabuck, K. Raghavachari, J. B. Foresman, J. V. Ortiz, Q. Cui, A. G. Baboul, S. Clifford, J. Cioslowski, B. B. Stefanov, G. Liu, A. Liashenko, P. Piskorz, I. Komaromi, R. L. Martin, D. J. Fox, T. Keith, M. A. Al-Laham, C. Y. Peng, A. Nanayakkara, M. Challacombe, P. M. W. Gill, B. Johnson, W. Chen, M. W. Wong, C. Gonzalez, and J. A. Pople, Gaussian, Inc., Wallingford CT, 2004

\bibitem{Lee89} T. J. Lee, J. E. Rice, G. E. Scuseria, and H. F. Schaefer III, {\em Theor. Chim. Acta} {\bf 75}, 81 (1989); T. J. Lee, and P. R. Taylor, {\em ibid} {\bf 23}, 199 (1989). 
\bibitem{Lei00} M. L. Leininger, I. M. B. Nielsen, T. D. Crawford, and C. L. Janssen, {\em Chem. Phys. Lett.} {\bf 328}, 431 (2000); I. M. B. Nielsen, and C. L. Janssen, {\em ibid.} {\bf 310}, 568 (1999); C. L. Janssen, and I. M. B. Nielsen, {\em ibid.} {\bf 290}, 423 (1998).
18:41

\bibitem{C2} C. D. Sherrill and P. Piecuch, {\em J. Chem. Phys.} {\bf 122}, 124104 (2005) and references therein.

\bibitem{C2Bartlett} S. A. Kucharski, M. Kolaski, and R. J. Bartlett, {\em J. Chem. Phys.} {\bf 114}, 692 (2001).

\bibitem{Feller} B. Ruscic, private communication [unpublished data based on Active Thermochemical Tables, ATcT Version 1.25 and the Core (Argonne) Thermochemical Network Version 1.056 2006], quoted in: D. Feller and K. A. Peterson, {\em J. Chem. Phys.} {\bf 126}, 114105 (2007).

\bibitem{Hub79} K. P. Huber and G. Herzberg, {\em Constants of diatomic molecules} (Van Nostrand Reinhold, New York, 1979).

\bibitem{BN} A. Karton and J. M. L. Martin, {\em J. Chem. Phys.} {\bf 120}, 144313 (2006).

\bibitem{Watts92}
J. D. Watts and R. J. Bartlett, {\em Chem. Phys. Lett.} {\bf 190}, 19 (1992).
\bibitem{Parasuk92}
V. Parasuk and J. Alml\"of, {\em Theor. Chim. Acta} {\bf 83}, 227 (1992).
\bibitem{QMC}
Y. Shlyakhter, S. Sokolova, A. L\"uchow, and J. B. Anderson, {\em J. Chem. Phys.} {\bf 110}, 10725 (1999).

\bibitem{SCS-MP2} S. Grimme, {\em J. Chem. Phys.} {\bf 118}, 9095 (2003). For a physical interpretation of the
SCS-MP2 method in terms of Feenberg scaling, see A. Szabados, {\em J. Chem. Phys.} {\bf 125}, 214105 (2006)

\bibitem{SCS-CCSD} T. Takatani, E. E. Hohenstein, and C. D. Sherrill, {\em J. Chem. Phys.} {\bf 128}, 124111 (2008). Those using quantum chemical codes such as MOLPRO that report $E_S$ and $E_T$ rather than $E_{\alpha\alpha}$, $E_{\beta\beta}$, and $E_{\alpha\beta}$ can easily obtain the SCS-CCSD correlation energy as 1.17666$E_T$+1.27$E_S$.


\bibitem{g2mp2} Curtiss, L. A.; Raghavachari, K.; and Pople, J. A. {\em J. Chem. Phys.} {\bf 98}, 1293 (1993)
\bibitem{g2} Curtiss, L. A.; Raghavachari, K.; Trucks, G. W.; and Pople, J. A. {\em J. Chem. Phys.} {\bf 94}, 7221 (1991)
\bibitem{g3} Curtiss, L. A.; Raghavachari, K.; Redfern, P. C.; Rassolov, V.; and Pople, J. A. {\em J. Chem Phys.} {\bf 109}, 7764 (1998)
\bibitem{g3b3} Baboul, A. G.; Curtiss, L. A.; Redfern, P. C.; and Raghavachari, K. {\em J. Chem. Phys.} {\bf 110}, 7650 (1999)
\bibitem{g4} Curtiss, L. A.;  Redfern, P. C.; and Raghavachari, K. {\em J. Chem. Phys.} {\bf 126}, 084108 (2007)
\bibitem{G4MP2} L. A. Curtiss, P. C. Redfern, and K. Raghavachari, {\it J. Chem. Phys.} {\bf 127}, 124105 (2007)
\bibitem{cbs-qb3} Montgomery Jr, J. A.; Frisch, M. J.; Ochterski, J. W.; and Petersson, G. A.; {\em J. Chem. Phys.} {\bf 110}, 2822 (1999)
\bibitem{cbs-apno} Ochterski, J. W.; Petersson, G. A.; and Montgomery Jr., J. A. {\em J. Chem. Phys.} {\bf 104}, 2598 (1996)

\bibitem{DFT_ref_val}C$_2$($^1\Sigma_g^+$) W4.4 theory; C$_2$($^3\Pi_u$) W4.3 theory; C$_3$--C$_5$ W4 theory; C$_6$($^3\Sigma_g^-$) W4 theory via the 2C$_4$($^3\Sigma_g^-$)$\rightarrow$C$_6$($^3\Sigma_g^-$)+C$_2$($^3\Pi_u$) reaction energy at the W3.2 level; C$_6$($^1A_1^\prime$) W3.2 theory; C$_7$($^1\Sigma_g^+$) W4 theory via the 2C$_5$($^1\Sigma_g^+$)$\rightarrow$C$_7$($^1\Sigma_g^+$)+C$_3$($^1\Sigma_g^+$) reaction energy at the W3.2lite level; C$_8$($^3\Sigma_g^-$) W3.2 theory via the 2C$_6$($^3\Sigma_g^-$)$\rightarrow$C$_8$($^3\Sigma_g^-$)+C$_4$($^3\Sigma_g^-$) reaction energy at the W3.2 level; C$_8$($^1A_g$) W3.2lite theory; C$_9$($^1\Sigma_g^+$) W4 theory via the 3C$_5$($^1\Sigma_g^+$)$\rightarrow$C$_9$($^1\Sigma_g^+$)+2C$_3$($^1\Sigma_g^+$) reaction energy at the W3.2lite level; C$_{10}$($^3\Sigma_g^-$) W3.2lite theory via the 2C$_6$($^3\Sigma_g^-$)$\rightarrow$C$_{10}$($^3\Sigma_g^-$)+C$_8$($^3\Sigma_g^-$) reaction energy at the W2.2 level; and C$_{10}$($^1A_1^\prime$) W2.2 theory.
\bibitem{HCTH407} Boese, A. D.; Handy, N. C. {\em J. Chem. Phys.} {\bf 114}, 5497 (2001)
\bibitem{B88ex} Becke, A. D. {\em Phys. Rev. A} {\bf 38}, 3098 (1988)
\bibitem{LYPc} Lee, C.; Yang, W.; Parr, R. G. {\em Phys. Rev. B} {\bf 37}, 785 (1988)
\bibitem{PW91c} Perdew, J. P.; Chevary, J. A.; Vosko, S. H.; Jackson, K. A.; Pederson, M. R.; Singh, D. J.; Fiolhais, C. {\em Phys. Rev. B} {\bf 46}, 6671 (1992)
\bibitem{P86c} Perdew, J. P. {\em Phys. Rev. B} {\bf 33}, 8822 (1986)
\bibitem{PBE} (a) Perdew, J. P.; Burke, K.; Ernzerhof, M. {\em Phys. Rev. Lett.} {\bf 77}, 3865 (1996); Erratum: {\bf 78}, 1396 (1997)
\bibitem{M06-L} Zhao, Y.; Truhlar, D. G. {\em J. Chem. Phys.} {\bf 125}, 194101 (2006)
\bibitem{VSXC} van Voorhis, T.; Scuseria, G. E. {\em J. Chem. Phys.} {\bf 109}, 400 (1998)
\bibitem{TPSS} Tao, J. M.; Perdew, J. P.; Staroverov, V. N.; Scuseria, G. E. {\em Phys. Rev. Lett.} {\bf 91}, 146401 (2003)
\bibitem{t-HCTH} Boese, A. D.; Handy, N. C. {\em J. Chem. Phys.} {\bf 116}, 9559 (2002)
\bibitem{PBE0} Adamo, C.; Barone, V. {\em J. Chem. Phys.} {\bf 110}, 6158 (1999)
\bibitem{B97-2} Wilson, P. J.; Bradley, T. J.; Tozer, D. J. {\em J. Chem. Phys.} {\bf 115}, 9233 (2001)
\bibitem{B3} Becke, A. D. {\em J. Chem. Phys.} {\bf 98}, 5648 (1993)
\bibitem{B97-1} Hamprecht, F. A.; Cohen, A. J.; Tozer, D. J.; Handy, N. C. {\em J. Chem. Phys.} {\bf 109}, 6264 (1998)
\bibitem{B98} Schmider, H. L.; Becke, A. D. {\em J. Chem. Phys.} {\bf 108}, 9624 (1998)
\bibitem{TPSSh} Staroverov, V. N.; Scuseria, G. E.; Tao, J.; Perdew, J. P. {\em J. Chem. Phys.} {\bf 119}, 12129 (2003)
\bibitem{B3LYP} Stephens, P. J.; Devlin, F. J.; Chabalowski, C. F.; Frisch, M. J. {\em J. Phys. Chem.} {\bf 98}, 11623 (1994)
\bibitem{mPW1K} Lynch, B. J.; Fast, P. L.; Harris, M.; Truhlar, D. G. {\em J. Phys. Chem. A} {\bf 104}, 4811 (2000)
\bibitem{mPWex} Adamo, C.; Barone, V. {\em J. Chem. Phys.} {\bf 108}, 664 (1998)
\bibitem{BHLYP} Becke, A. D. {\em J. Chem. Phys.} {\bf 98}, 1372 (1993)
\bibitem{mPW1B95} Zhao. Y.; Truhlar, D. G. {\em J. Phys. Chem. A} {\bf 108}, 6908 (2004)
\bibitem{B95c} Becke, A. D. {\em J. Chem. Phys.} {\bf 104}, 1040 (1996)
\bibitem{M06} (a) Zhao, Y.; Truhlar, D. G. {\em Theor. Chem. Acc.} {\bf 120}, 215 (2007); (b) Zhao, Y.; Truhlar, D. G. {\em Acc. Chem. Res.} {\bf 41}, 157 (2008)
\bibitem{PW6B95} Zhao, Y.; Truhlar, D. G. {\em J. Phys. Chem. A} {\bf 109}, 5656 (2005)
\bibitem{TPSS1KCIS} Zhao, Y.; Lynch, B. J.; Truhlar, D. G. {\em Phys. Chem. Chem. Phys.} {\bf 7}, 43 (2005)
\bibitem{M05} Zhao, Y.; Schultz, N. E.; Truhlar, D. G. {\em J. Chem. Phys.} {\bf 123}, 161103 (2005)
\bibitem{BMK} Boese, A. D.; Martin, J. M. L. {\em J. Chem. Phys.} {\bf 121}, 3405 (2004)
\bibitem{BB1K} Zhao, Y.; Lynch, B. J.; Truhlar, D. G. {\em J. Phys. Chem. A} {\bf 108}, 2715 (2004)
\bibitem{B2-PLYP} Grimme, S. {\em J. Chem. Phys.} {\bf 124}, 034108 (2006)
\bibitem{mPW2-PLYP} Schwabe, T.; Grimme, S. {\em Phys. Chem. Chem. Phys.} {\bf 8}, 4398 (2006)
\bibitem{B2K-PLYP} Tarnopolsky, A.; Karton, A.; Sertchook, R.; Vuzman, D.; Martin, J. M. L. {\em J. Phys. Chem. A} {\bf 112}, 3 (2008)
\bibitem{B2GP-PLYP} Karton, A.; Tarnopolsky, A.; Lam\`ere, J.-F.; Schatz, G. C.; Martin, J. M. L. {\em J. Phys. Chem. A}
{\bf 112}, 12868 (2008).

\bibitem{pc-2} Jensen, F. {\em J. Chem. Phys.} {\bf 115}, 9113 (2001); erratum {\bf 116}, 3502 (2002)

\bibitem{Francois} Giuffreda, M. G.; Deleuze, M. S.; Fran\c{c}ois, J.-P. {\em J. Phys. Chem. A} {\bf 2002}, {\it 106}, 8569.

\bibitem{Mar96c11} J. M. L. Martin, J. El-Yazal, and J.-P. Fran\c{c}ois, {\em Chem. Phys. Lett.} {\bf 252}, 9 (1996).


\bibitem{Petersson} 
(a) Petersson, G. A.; Tensfeldt, T.; Montgomery, J. A. J. {\em Chem. Phys.} {\bf 94}, 6091 (1991); 
(b) Montgomery, J. A.; Ochterski, J. W.; Petersson, G. A. {\em J. Chem. Phys.} {\bf 101}, 5900 (1994);
(c) Petersson, G. A. Personal communication

\end{thebibliography}
\end{document}